\documentclass[12pt]{article}

\usepackage{amsmath,amssymb,amsthm,eurosym,lineno,mathrsfs,color}
\usepackage[hyphens]{url}
\usepackage[colorlinks=true,linkcolor=blue,citecolor=blue,urlcolor=blue]{hyperref}
\usepackage{graphicx}

\numberwithin{equation}{section}

\theoremstyle{plain}
\newtheorem{theorem}{Theorem}
\newtheorem{lemma}[theorem]{Lemma}

\def\R {\mathbb{R}}

\def\Var{{\rm Var}}

\begin{document}


\title{\bf\large Constant Proportion Debt Obligations, Zeno's Paradox, \\
and the Spectacular Financial Crisis of 2008}

\author{{Donald Richards}\thanks{Department of Statistics, 
Penn State University, University Park, PA 16802.}
\ \ {and \ Hein Hundal}\thanks{Applied Research Laboratory, Penn State University, Science Park Road, State College, PA 16801.
\endgraf
\ {\it 2000 Mathematics Subject Classification}: Primary 91G20; 
Secondary 60K30.
\endgraf
\ {\it Key words and phrases}.  AAA-rating; Binomial coefficient; Binomial distribution; Cash-In event; Cash-Out event; Central Limit Theorem; Collateralized debt obligation (CDO); Credit default swap; Credit derivative product company; Exotic synthetic CDO; Financial derivatives; Financial ``engineering''; Gambler's fallacy; Geometric series; Golden Ratio; Law of Total Probability; LIBOR; Martingale; Mean-reversion; Pascal's triangle; Ratings agency; Special purpose vehicle; Structured finance.}
}

\date{June 13, 2012}
\maketitle

\begin{abstract}
\footnotesize{
We analyze a coin-tossing model used by a ratings agency to justify the invention of constant proportion debt obligations (CPDOs), and prove that it was impossible for CPDOs to achieve in a finite lifetime the Cash-In event of doubling its capital.  In the best-case scenario in which the coin is two-headed, we show that the goal of attaining the Cash-In event in a finite lifetime is precisely the goal, described more than two thousand years ago in Zeno's Paradox of the Dichotomy, of obtaining the sum of an infinite geometric series with only a finite number of terms.  In the worst-case scenario in which the coin is two-tailed, we prove that the Cash-Out event occurs in exactly ten tosses.  

For the case in which the coin is fair, we show that if a CPDO were allowed to toss the coin without regard for the Cash-Out rule then, eventually, the CPDO has a high probability of attaining high net capital levels; however, hundreds of thousands of tosses may be necessary to do so.  Moreover, if after a large number of tosses the CPDO shows a loss then the probability is high that it will Cash-Out on the very next toss.  In the case of a CPDO that experiences a tail on the first toss, or on any early toss, we show that, with high probability, the CPDO will have capital losses thereafter for hundreds of tosses, and we show that the sequence of net capital levels is a martingale.  When the Cash-Out rule is in force, we modify the Cash-In event to mean that the CPDO attains a profit of 90\% on its initial capital; then, we prove that the CPDO game, almost surely, will end in a finite number of tosses and the probability of Cash-Out is at least 89\%.

In light of these results, our fears about the durability of the on-going worldwide financial crisis are heightened by the existence of other financial derivatives more arcane than CPDOs.  In particular, we view askance all later-generation CPDOs that depend on an assumption of mean-reversion or utilize a betting strategy similar to that of their first-generation counterparts.

\vskip 0.5truein
\phantom{a}
{ }}
\end{abstract}

\parskip=2.5pt

\section{Introduction}
\label{intro}
\setcounter{equation}{0}

{\parskip=-1.5pt
{\sl 
\hfill{\small The tale is as old as the Eden Tree -- and new as the new-cut tooth --}

\hfill{\small For each man knows ere his lip-thatch grows he is master of Art and Truth;}

\hfill{\small And each man hears as the twilight nears, to the beat of his dying heart,}

\hfill{\small The Devil drum on the darkened pane: ``You did it, but was it Art?''}

\smallskip

\hfill{\small -- Rudyard Kipling, {\it The Conundrum of the Workshops}}
}
}

\bigskip
\bigskip

Constant proportion collateralized debt obligations (CPDOs) were designed in 2006 as a new product in the ``synthetic'' collateralized debt obligation market.  CPDOs were developed by financial engineers (known colloquially as ``quants'') at ABN Amro Securities and first sold to the public in late 2006 with fanfare and AAA ratings \cite{davies}.  

To create a CPDO, an investment bank forms a subsidiary corporation, called a ``special purpose vehicle'' (SPV).  The SPV sells debt, called CPDO notes,  to ``investors'' and deposits the proceeds in highly-rated collateral securities, such as government or high-grade commercial bonds, earning interest at a supposedly risk-free rate.  The SPV next uses the risk-free account as collateral to make a highly-leveraged sale of protection on a widely-traded, investment-grade index of credit default swaps, such as the iTraxx Europe or Dow Jones CDX; at regular six-month intervals, the sale is wound down, any profits from the sale are paid into the deposit-account and any losses are paid out of the account.  The SPV then repeats the process with a new highly leveraged sale of protection on the index, and the leverage factor is adjusted so that the SPV's obligations can be met from the deposit-account.  In general, the level of leveraging depends on the difference between the current value of the SPV's future liabilities and its net asset value \cite[Section 1]{gordywilleman}.  

CPDOs were designed to attract AAA ratings while paying a coupon rate substantially higher than similarly-rated instruments and without added risk.  In particular, ABN Amro's SURF notes supposedly represented AAA-rated debt, paying coupon rates 2\% higher than the London interbank offered rate, or LIBOR, and with no greater risk.  Because of these purported properties, CPDOs received widespread acclaim.  

From the start, CPDOs were hailed as innovative.  ABN Amro's SURF was awarded the ``Deal of the Year'' prize for 2006 by {\it Risk Magazine}.  The {\it International Financial Law Review}, in giving SURF its ``Award of 2006'', described it as the ``Innovation of the Year ... [the] ultimate recognition for achievement in global capital markets.''  SURF also received from {\it Euromoney} magazine the accolade of ``one of six `Deals of the Year 2006'\,'' \cite[p. 60, Appendix D]{bis}.  Undoubtedly, the highest praise for SURF came from Bear, Stearns, a well-known investment bank, where analysts waxed religious in praising CPDOs as the ``Holy Grail of structured finance'' \cite{jones3}.  Thus, by early 2007, estimates of CPDOs outstanding ranged as high as \EUR5.2 billion \cite[p. 62]{bis}, although CPDOs did not receive outright approval from prominent financial organizations; see \cite[Annex 1.3, p. 51]{imf}, \cite[pp. 5, 22--23, 85--86]{ecb}, and \cite[p. 199]{boe}.

And yet, CPDOs were problematic from inception.  In 2007, Fitch Ratings and Dominion Bond Rating Service, and others criticized the structure of CPDOs \cite{glover}.  In early 2008, the Bank for International Settlements \cite[pp. 39, 60--72]{bis} critiqued the nature of CPDOs, and concerns were raised by the U.S. Securities and Exchange Commission and members of Congress.  Moody's Investors Service subsequently acknowledged errors in its CPDO ratings, attributed the difficulties to computer software errors, and downgraded many CPDOs \cite{jones2}.  Finally, by late 2008, many CPDOs had defaulted \cite{khasawneh}.

That a CPDO could be AAA-rated and provide significantly higher coupon rates without greater risk seemed to us to be unrealistic and self-contradictory.  Such a debt instrument surely had to be a ``free lunch,'' a phenomenon whose existence contradicted the efficient-market hypothesis, a key tenet of financial engineering.  Thus, it would be curious if the quants could devise an instrument whose existence contradicted a basic axiom of their field.  Moreover, if it were possible to devise AAA-rated CPDOs that pay a risk-free 2\% rate above LIBOR then it should be possible to devise AAA-rated (CPDOs)$^2$ that pay a risk-free 2\% above CPDOs, then AAA-rated (CPDOs)$^3$ that pay a risk-free 2\% above (CPDOs)$^2$, {\it ad infinitum}, a process that clearly is implausible.  

We were also intrigued by the remarkable, although not unprecedented, speed with which these novel financial entities floundered.  In seeking to understand the basic reason for their rapid descent into insolvency, our curiosity was heightened by a news article \cite{wsjcpdo} whose last paragraph read as follows:

\begin{quote}
``In a 2006 `primer' on CPDOs written by a team of authors [...], 
Moody's used an analogy that compared the product to a `coin-toss game.'  The strategy `is based on the notion that if `heads' has appeared more frequently than expected, it is less likely to continue appearing,' the report said.''
\end{quote}

In the study of random phenomena, such as the repeated tossing of a physical coin, it is well-known that the Law of Averages \cite[p. 274]{fpp} contradicts the ``notion that if `heads' has appeared more frequently than expected, it is less likely to continue appearing.''  Indeed, that notion is precisely the concept of {\it gambler's fallacy} \cite{freedman97,tversky}.  Consequently, we decided to analyze the probabilistic nature of CPDOs and, as we shall prove by means of elementary probability theory, there are serious flaws with the basic structure of CPDOs defined by such a ``coin-toss game.''  As we investigate the probabilistic behavior of such a game, what we shall discover is well-expressed by a comment made in another context by David Freedman \cite[p. 102]{freedman87}:  

\begin{quote}
``At bottom, my critique is pretty simple-minded: Nobody pays much 
\hfill\break
attention to the assumptions, and the technology tends to overwhelm 
\hfill\break
commonsense.''
\end{quote}

In this paper, we shall use undergraduate-level methods to study the coin-tossing model for CPDOs provided in \cite{cpdoprimer}.  We begin by proving that, even if the coin is two-headed, the CPDO cannot achieve the Cash-In event of doubling the initial capital within a finite lifetime, and we relate the case of the two-headed coin to a classical paradox of Zeno, the Greek philosopher.  If the coin is two-tailed, on the other hand, we prove that the Cash-Out event occurs in exactly ten tosses.  

We consider the case in which the coin is fair, i.e., on any toss, heads and tails have equal probability of occurring; however, all our calculations can be easily modified to cover the case of an unfair coin.  We show that if a CPDO is allowed to toss the coin without regard to losses then, eventually, it has a high probability of amassing large net capital; however, hundreds of thousands of tosses may be necessary to do so.  On the other hand, if, after a large number of tosses the CPDO shows a loss then, with high probability, it will Cash-Out on the very next toss.  In the case of a CPDO that experiences a tail on the first or on an early toss, we show that, with high probability, the CPDO will have capital losses after hundreds of additional tosses, and we obtain explicit formulas for those probabilities.  More generally, we show that the CPDO betting strategy does not have lack-of-memory: If at any given stage the CPDO shows a loss then, at all future stages, the probability of a loss, or of Cash-Out, generally increases.  Further, we show that the sequence of net capital levels is a martingale.  

As we noted earlier, it is impossible for the CPDO to double its initial capital within a finite lifespan.  Thus, for the case in which the CPDO is subject to the Cash-Out rule, we modify the Cash-In event to mean that the CPDO attains a profit of 90\% on its initial capital.  Then, we prove that the CPDO game, almost surely, will end in a finite number of tosses.  Further, we prove that the probability that a CPDO will Cash-Out is at least 89\%, and we provide in Section \ref{simulations} figures that illustrate this phenomenon.  

On the basis of these results, we conclude that a CPDO was little more than a recipe for continual losses over lengthy periods, and with a high probability of default.  Moreover, second-generation CPDOs \cite{ubscpdoflaws} are likely to suffer the same fate if they employ betting strategies similar to those of their first-generation counterparts.


\section{A ratings agency's coin-toss model for CPDOs}
\label{cointossmodel}
\setcounter{equation}{0}

We provide in this section a quotation taken from \cite[pp. 3--4] {cpdoprimer}, a publication which came to our attention because of the newspaper article \cite{wsjcpdo} (see also \cite{dorn,katsaros}):

\begin{quote}
``{\bf An Analogy}

``To understand how the CPDO works, we can draw a parallel with a simple coin-toss game. The outcome of the toss is either `heads' or `tails'. Heads results in a 100\% return and tails a 100\% loss.

``The player has an initial stake of 1000, comprised of 100 from his own pocket and 900 borrowed from a friend.  At the outset his strategy is as follows: if he succeeds in converting his stake into 2000 of winnings, he will stop and reimburse his friend, having thus multiplied his initial investment by 11.  This corresponds to the Cash-In Event.

``At the same time, his friend is concerned about his stake and thus if the player loses more than 100, he will stop playing.  This corresponds to the Cash-Out Event. (For each toss, the player bets 1\% of the difference between his current stake and 2000 -- the simple rebalancing rule in our example.)

``If he bets 10 from the initial stake on `heads', there are two possible outcomes:

``-- `Heads': the player's stake rises to 1010, so at the next round he will bet only 9.90.

``-- `Tails': he now has only 990, so at the next round he will bet 10.10.

``Such a strategy is based on the notion that if `heads' has appeared more frequently than expected, it is less likely to continue appearing, and similarly for `tails.'  In other words, the strategy is based on the concept of `mean-reversion.'\,''
\end{quote}

\section{Remarks on the coin-toss model}
\label{cpdos}
\setcounter{equation}{0}

There are several problems with the strategy described in Section \ref{cointossmodel}.  We deal with them in a number of subsections, as follows:

\subsection{CPDO: Friend or foe?}
\label{friendorfoe}
\setcounter{equation}{0}

It is striking that the word ``friend'' is used to describe the entity from which ``900'' is ``borrowed'', especially so when, in the real world, ``an initial stake of 1000'' corresponds to hundreds of millions of U.S. dollars.  It is plausible that the use of the word ``friend'' may have encouraged among CPDO participants a subconscious lowering-of-the-guard as to possible negative outcomes of the ``game.''  

In our view, CPDO participants would have been better served had the word ``banker'' been substituted for ``friend.''  Indeed, to judge by the way in which banks, broker-dealers, and hedge funds co-exist in the highly competitive financial world, it would have been better 
had the word ``friend'' been replaced by ``foe'' throughout \cite[pp. 3--4]{cpdoprimer}.  (Here, we note that some hedge funds have complained that their bankers or broker-dealers were overly abrupt in closing margin accounts after a short string of ``tails,'' thereby forcing the funds into insolvency.)

Similar remarks apply to the use of the words ``player'' and ``game.''  Perhaps ``battle'' would have been a more realistic term than ``game''; similarly, substitute ``warrior,'' ``gambler,'' or ``speculator'' for ``player.''  

In case a reader views our concerns about the language of the quotation as an over-occupation with mere semantics, we note that analysts at an investment bank also posed the question, ``CPDO: Friend or foe?'' \cite[p. 572]{ms}.

\subsection{Mean-reversion and gambler's fallacy}
\label{meanreversion}
\setcounter{equation}{0}

We have greater concerns with the ``concept of `mean-reversion' '' referred to in \cite{cpdoprimer}.  

As we observed before, in the repeated tossing of a physical coin, the ``notion that if `heads' has appeared more frequently than expected, it is less likely to continue appearing'' is precisely the concept of {\it gambler's fallacy}, the misbelief that the long-run relative frequency of a random event will apply even in the short run.  Classic examples of gambler's fallacy are well-known in casino gambling, where players often believe that a run of good luck is highly likely to follow a run of bad luck.  Because individual outcomes of the tosses of a coin are independent random events, however, a long string of bad luck may well be followed by another long string of the same.  (Indeed, it is well-known in probability theory that, in the indefinite tossing of a fair coin, the probability that a given string of heads or tails will occur infinitely often is 100\% \cite[p. 61]{billingsley}.)

Tversky and Kahneman \cite[p. 106]{tversky} describe gambler's fallacy as the phenomenon wherein 
\begin{quote}
``$\ldots$ subjects [in a probability learning experiment] act as if {\it every} segment of the random sequence must reflect the true proportion: if the sequence has strayed from the population proportion, a corrective bias in the other direction is expected. $\ldots$  [The]  heart of the gambler's fallacy is a misconception of the fairness of the laws of chance.  The gambler feels that the fairness of the coin entitles him to expect that deviation in one direction will soon be cancelled by a corresponding deviation in the other.  Even the fairest of coins, however, given the limitations of its memory and mortal sense, cannot be as fair as the gambler expects it to be.''
\end{quote}

\noindent
Freedman \cite{freedman97} has described connections between misconceptions in elementary probability theory and gambler's fallacy, which he describes as ``a cognitive illusion whose power is demonstrated by Cohen'' \cite{cohen}.  

Although mean-reversion may have been observed in the past behavior of a financial time series, it is not clear to us how it can be predicted that future observations of the same time series will continue to exhibit mean-reversion.  We sense an overuse of the concept among financial analysts in the same way that na\"ive casino gamblers hope for mean-reversion away from runs of bad luck (but never from runs of good luck).

\subsection{The CPDO as sub-optimal betting strategy}
\label{optimal}
\setcounter{equation}{0}

The betting strategy in the CPDO coin-toss model is curious in that it leads to a larger bet after a loss and a smaller bet after a win.  Such a strategy is opposite to Kelly's formula 
\cite{kelly,thorp} and to Whitworth's formula \cite{sr1,sr2,whitworth}, for those two formulas reduce the size of the bet following a loss and increase it after a win.  The Kelly and Whitworth formulas are well-known to satisfy optimality properties with regard to maximization of expected logarithmic utility or conservation of capital \cite{rump,thorp}, so the CPDO strategy is unlikely to enjoy similar properties.

The CPDO strategy differs from standard methods in the theory of gambling systems \cite{billingsley,dubinssavage,maitrasudderth}.  The strategy is not {\it timid} (i.e, betting a fixed amount on each toss), which may be commendable given the dangers of timidity \cite[p. 108]{billingsley}.  Also, the strategy is not {\it bold} (i.e., betting either the amount needed to double the stake in one win, or the total remaining stake, whichever is smaller), which may be unwise given that boldness is optimal in the sense that it maximizes the probability of attaining Cash-In \cite{billingsley,maitrasudderth}.  We also recall that, regardless of strategy, the probability of doubling the stake in a subfair casino is at most 50\% \cite[p. 1]{dubinssavage}; thus, after borrowing 900 from a ``friend,'' and ignoring all costs of operations, a real-world CPDO had {\it at most} a 50\% chance of attaining the Cash-In event.  In the case of timid play, bounds on the probability of Cash-In can be obtained by means of Chernoff's inequality \cite[Example 5f, pp. 452--453]{ross}.

\subsection{Zeno's paradox and the impossibility of Cash-In}
\label{zeno}
\setcounter{equation}{0}

It will be simple to prove the following result.

\begin {theorem} 
\label{bestworst}
In the best-case scenario in which the coin is two-headed, it is impossible for the CPDO to double its stake in a finite number of tosses.  In the worst-case scenario in which the coin is two-tailed, the CPDO will Cash-Out in exactly $10$ tosses.
\end {theorem}

In the case of a two-headed coin, a CPDO will be ensnared in {\sl The Dichotomy} \cite{zeno}, a paradox of motion formulated by Zeno of Elea, the Greek philosopher (b. $\sim$ 490 BC).  In such a paradox the mythical Atalanta, a fast runner, runs 1 mile in the first minute, 1/2 mile in the next minute, 1/4 mile in the next minute, and so on, in perpetuity.  Atalanta would require an infinite amount of time to run two miles, for the distances covered in successive minutes follow a geometric progression whose sum is less than $2$ for any finite number of minutes.  In the case of a CPDO equipped with a two-headed coin, we shall prove that its stake increases with each toss by successively smaller increments that form a geometric progression; hence, similar to Atalanta, the CPDO will require an infinite number of tosses to double its starting capital.

Given the enormous amounts of time and financial resources committed to real-world CPDOs, and the enormity of any problems that would arise from flaws in the design of CPDOs, it is curious that the connection with Zeno's paradox was overlooked entirely by CPDO inventors, purchasers, ratings agencies, and financial regulators.  
\,\footnote{Admittedly, we ourselves overlooked it, and we are grateful to \href{http://www.stat.lsa.umich.edu/~moulib/}{Moulinath Banerjee}, of the University of Michigan, for pointing out the connection to us.  We can, at least, offer the excuses that we are not quants, regulators, or ratings analysts, and in comparison with real-world CPDO operators, we have scant financial resources to conduct full-time research on the subject.}

\section{Probabilistic analysis of the CPDO model}
\label{probability}
\setcounter{equation}{0}

Let us turn now to a probabilistic analysis of the coin-toss CPDO model in \cite{cpdoprimer}.  Although such an analysis can be done using the advanced methods of Bilingsley \cite[Section 7]{billingsley}, Dubins and Savage \cite{dubinssavage}, or Maitra and Sudderth \cite{maitrasudderth}, we will utilize only elementary probabilistic concepts familiar to {\it undergraduate} students, generally referring to the undergraduate-level textbook \cite{ross}.  Admittedly, we want to underscore that our concerns about the CPDO model could have been formulated by a well-trained undergraduate student, not to mention world-class experts in financial engineering.  

Let $C_1, C_2, C_3,\ldots$ represent the outcomes of successive coin 
tosses, where 
\begin{equation}
\label{cj}
C_j = \begin{cases}
+1, & \hbox{if the $j$th toss results in heads } \\
-1, & \hbox{if the $j$th toss results in tails.}
\end{cases}
\end{equation}
Throughout the paper, the constants 
\begin{equation}
\label{gamma}
\gamma = 0.1,
\end{equation}
and 
\begin{equation}
\label{delta}
\delta = 0.01,
\end{equation}
will play a crucial role, and we display these equations to emphasize their repeated use.  Readers may change the values of these constants to suit their individual needs, and the calculations appearing in the paper can be modified in a straightforward manner.

For $k = 0,1,2,\ldots,$ denote by $Y_k$ the size of the stake held by the CPDO on the $k$th toss.  We take $Y_0$, the size of the initial stake, to be $1$ unit, consisting of $\gamma = 0.1$ units of personal capital and $1-\gamma = 0.9$ units of borrowed capital.  (To retain a connection with real-world CPDO practices, we encourage the reader to view each ``unit'' as representing the small sum of one hundred million U.S. dollars.)  

We define the Cash-Out event more conservatively than in the coin-toss model in Section \ref{cointossmodel}, adopting the rule that the CPDO will Cash-Out if the capital falls {\it to or below} $\gamma$, the cut-off level.  Hence, if the CPDO's capital falls to $\gamma$, the bank will Cash-Out the account rather than permit another coin toss and risk any loss on its loan.  

From the strategy of betting the proportion $\delta$, of the difference between the goal of $2$ and the current stake, it follows that $Y_1,Y_2,Y_3,\ldots$ satisfy the recurrence relation 
\begin{equation}
\label{yrecurrence}
Y_k = Y_{k-1} + \delta (2-Y_{k-1})C_k \, ,
\end{equation}
where $k = 1,2,3,\ldots$.  To simplify the algebraic calculations in the ensuing analysis, it is convenient to work with $X_k = Y_k - 1$, the CPDO's net profit (or \textcolor{red}{loss}) at the $k$th stage.  It is easy to see from (\ref{yrecurrence}) that $X_0 = 0$ and, for $k=1,2,3,\ldots$, 
\begin{equation}
\label{xkrecurrence}
X_k = X_{k-1} + \delta(1-X_{k-1})C_k.
\end{equation}   
It is an undergraduate-level exercise to solve the recurrence relation (\ref{xkrecurrence}), and then we obtain the following explicit formula for $X_k$.  

\medskip

\begin {theorem}
\label{xkformula}
For $k = 0,1,2,\ldots$, 
\begin{equation}
\label{xkexplicit}
X_k = 1 - \prod_{j=1}^k (1-\delta C_j).
\end{equation}
\end {theorem}

\bigskip 

We can now verify the best- and worst-case scenarios described in Theorem \ref{bestworst}.  

\bigskip 

\noindent
{\it Proof of Theorem \ref{bestworst}}.  
If the coin is two-headed then $C_k = +1$ for all $k$ and then, by (\ref{xkexplicit}), we have $X_k = 1 - (1-\delta)^k$.  Since $\delta < 1$ then $0 < (1-\delta)^k < 1$, therefore $0 < X_k < 1$ for all $k$, and this proves that Cash-In cannot occur in a finite number of tosses.  

If the coin is two-tailed then $C_k = -1$ for all $k$; hence, by (\ref{xkexplicit}), $X_k = 1 - (1+\delta)^k$, and $X_k$ decreases as $k$ increases.  Solving for $k$ the equation $1 - (1+\delta)^k = -\gamma$, we obtain $k = \ln(1+\gamma)/\ln(1+\delta) = 9.58$; hence, Cash-Out occurs at the tenth toss.
$\qed$

\bigskip 

Henceforth, we assume in our analysis that the coins referred to in \cite{cpdoprimer} are fair, i.e., that the probability of heads is 50\%, and that individual coin tosses are mutually independent, so that each toss provides no information whatsoever about the outcome of other tosses.  In probabilistic terminology, we assume that the random variables $C_1, C_2, C_3,\ldots$ are mutually independent, and that, for all $j = 1,2,3,\ldots$, 
$$
P(C_j = +1) = P(C_j = -1) = 1/2.
$$
Then, each $C_j$ has a Bernoulli distribution with probability of success $1/2$.  

A real-world CPDO manager may argue that, in practice, neither assumption of a fair coin nor mutual independence of tosses is valid.  While such objections may be valid, Theorem \ref{bestworst} shows that neither assumption is crucial to the key point that the CPDO model cannot double the initial stake in a finite life time.  Further, if a protracted serial dependence between tosses were to hold in the real world then it will be noticed eventually, and will be exploited by the CPDO's opponents.  Therefore, it would be better for the CPDO that tosses be mutually independent.  

As regards the coin being fair, a real-world CPDO manager who assumes that his coin has greater than 50\% probability of heads surely is suffering from hubris.  By definition of ``probability'' as {\it long-term} relative frequency, he would be assuming that, on average, he will outperform a majority of market participants {\it in perpetuity}, and that those opponents will continue to compete with him despite his noticeable advantage.  On the other hand, if his coin is such that the probability of heads is less than 50\% then there is no financial advantage to tossing given that, on each toss, his chances of losing are greater than that of winning.  Indeed, if the probability of heads is less than 50\% then, almost surely, the CPDO will Cash-Out in finitely many tosses.  

In any case, the ensuing analysis can be easily extended to apply to any choice of the probability of heads.

We now have the following results for the mean, $E(X_k)$, and variance $\Var(X_k)$, of the sequence of net capital levels.  

\bigskip

\begin {theorem}
\label{expectxk}
For $k = 1,2,3,\ldots$, $E(X_k) = 0$ and 
\begin{equation}
\label{variance}
\Var(X_k) = (1+\delta^2)^k-1.
\end{equation}
In particular, $\Var(X_k)$ increases exponentially with $k$.  
\end {theorem}

\bigskip

These results are ominous for a CPDO manager who is more eager to avoid Cash-Out than to achieve Cash-In.  The initial net capital of $0$ is much closer to $-\gamma$, the Cash-Out level, than to $1$ so it is inauspicious that $\Var(X_k)$ grows exponentially with $k$ while, at the same time, $E(X_k) = 0$ for all $k$.  Moreover, that growth remains exponential in $k$, {\it regardless of the size of} $\delta$; so, even a more cautious CPDO which bets at each stage $\delta/2$, i.e., 50 basis points, of the difference between $X_k$ and the goal of 1, also will have exponential growth in $\Var(X_k)$ while having $E(X_k) = 0$ for all $k$.  

These observations alone would cause us to decline to commit billions of dollars to the CPDO coin-toss model.

\subsection{Cash-In: The case of the patient banker}
\label{patientbanker}

Let us now consider the situation in which the CPDO's banker is patient and will allow the CPDO to toss the coin indefinitely without regard for the Cash-Out rule.

Suppose that $k$, the number of tosses, is very large.  Because the coin is fair, the $C_j$ in (\ref{cj}) will be approximately equally divided between $+1$ and $-1$.  Then, by (\ref{xkexplicit}), 
$$
X_k \simeq 1 - (1-\delta)^{k/2}(1+\delta)^{k/2} = 1 - (1-\delta^2)^{k/2}.
$$
Intuitively, for large values of $k$, $X_k > 0$ and $X_k \to 1$ as $k \to \infty$.  Thus, if the banker is sufficiently patient then, with high probability, the CPDO eventually will achieve near-Cash-In, in particular, will attain positive net capital levels.  However, in making this statement more precise, we discover bad news hidden within this good news.  

\bigskip 

\begin {theorem}
\label{losses}
For any $t \in (0,1)$, 
\begin{equation}
\label{eventualloss}
\lim_{k \to \infty} P(X_k \ge t) = 1.
\end{equation}
However, 
\begin{equation}
\label{successivecashout}
\lim_{k \to \infty} P(X_{k+1} \le -\gamma \, | \, X_k < 0) = 1.
\end{equation}
\end {theorem}

\bigskip 

The proof of this result is provided in Section \ref{proofs}.  

The result (\ref{eventualloss}) shows that, eventually, the CPDO will, with high probability, attain a net capital level greater than any specified level $t$.  Thus, a patient banker is highly unlikely to have a loss on the CPDO loan.  And yet, the banker must be prepared to be extremely patient for the CPDO to amass high net capital; for instance, if the CPDO desires a 95\% probability in order that $X_k \ge 1-\gamma = 0.9$ then we shall show that about 108,220 tosses will be necessary.  Bearing in mind that real-world CPDO coin tosses were intended originally to occur at six-month intervals, 108,220 tosses represents over 54,000 years, a length of time which would stretch 
the patience of most bankers.  

Moreover, 
(\ref{successivecashout}) means that after a large number of tosses have occurred, if the CPDO shows a loss at the $k$th toss then the probability is high that it will Cash-Out on the next toss.  

Because $P(X_k \le -\gamma) \le P(X_k < 0)$, it follows from (\ref{eventualloss}) that 
$$
\lim_{k\to\infty} P(X_k \le -\gamma) = 0,
$$
i.e., the probability of Cash-Out eventually approaches 0.  Nevertheless, the CPDO should take little comfort from this result.  Indeed, let 
$$
\mu = \tfrac12\ln(1-\delta^2),
\qquad
\sigma^2 = \tfrac14\Big(\ln\frac{1+\delta}{1-\delta}\Big)^2,
$$
and denote by $\Phi(\cdot)$ the cumulative distribution function of the standard normal distribution; then, we shall prove later that, for large $k$, 
\begin{equation}
\label{approx}
P(X_k \le -\gamma) \simeq 
\Phi\Big(\frac{-\ln(1+\gamma) + k\mu}{\sqrt{k\sigma^2}}\Big).
\end{equation}
On calculating the right-hand side of (\ref{approx}) for $k \ge 25$, we find that it increases steadily from 2.7\% (at $k=25$) to 33.1\% (at $k = 1,906$), and then decreases steadily.  Hence, the probability of Cash-Out increases for many tosses before it decreases.  

The CPDO game clearly is not to be played for moderately long periods of time.

\subsection{The martingale connection}
\label{secmartingale}

\medskip

{\hfill\small{\sl ``Beware of such sirens, young man!  ... above all avoid a martingale, ...''}} 

\smallskip

{\hfill\small{-- W. M. Thackeray, {\sl The Newcomes: Memoirs Of A Most Respectable Family}}} 

\bigskip
\bigskip

The sequence $X_k$, $k=1,2,3,\ldots$ turns out to have a property even more dangerous than those described in Theorem \ref{losses}.  

\medskip

\begin {theorem}
\label{martingale}
The sequence $X_1,X_2,X_3,\ldots$ is a martingale: 
$E(X_k \, | \, X_1,\ldots,X_{k-1}) = X_{k-1}$,
almost surely, for all $k \ge 1$.
\end {theorem}

\medskip

This result is, in fact, a special case of an example given by Williams \cite[p. 95, Example (b)]{williams}, who shows that if $M_0 = 1$ and, for $k \ge 1$, $M_k$ is the product of $k$ independent, non-negative random variables, each with mean $1$, then $M_k$ is a martingale.  The martingale nature of $M_k$ appears to have been first recognized by Abraham Wald \cite{wald} in his fundamental work in statistical inference.  

The connection between CPDOs and martingales was also noted by Jones \cite{jones1} and Katsaros \cite{katsaros}, albeit in the layman's sense.  Still, it is well-known in the popular literature that martingale strategies are dangerous over protracted periods of time.  

The martingale property in Theorem \ref{martingale} entails that, for a CPDO which has observed the outcomes $X_1,\ldots,X_{k-1}$, the expected outcome of the next toss is $X_{k-1}$, the {\it last observed outcome}.  Such a result should be disquieting to the CPDO because, once it registers a net capital loss then, on average, the outcome of the next toss also is likely to result in negative net capital.  The martingale property does imply that if $X_{k-1} > 0$ then it can be expected that $X_k > 0$ also, but the danger is that the initial stake of $0$ net capital is much closer to $-\gamma$, the Cash-Out level, than it is to $1$, the Cash-In level.  

Therefore, {\it unless a CPDO expects a long run of heads from the outset}, the martingale property of $X_k$ suggests that the betting strategy will be troublesome.  To quantify this statement, we now calculate the probabilities of net losses for small values of $k$.  

\smallskip

On the first toss, the CPDO will have a net capital loss if and only if the toss results in tails; hence, $P(X_1 < 0) = 1/2$.  
At the second toss, it follows from (\ref{xkexplicit}) that $X_2 < 0$ if and only if $C_1 = C_2 = -1$, hence $P(X_2 < 0) = 1/4$.  
As for $P(X_3 < 0)$, by applying (\ref{xkexplicit}) once more, we find that $X_3 < 0$ if and only if at least two of $C_1$, $C_2$, and $C_3$ are equal to $-1$; therefore $P(X_3 < 0) = 1/2$.  
Also, by similar arguments, we obtain $P(X_4 < 0) = 5/16$ and $P(X_5 < 0) = 1/2$.  


\begin{table}[!h]
\caption{Probability of a Net Loss on the $k$th Toss}
\begin{center}
\renewcommand{\arraystretch}{1.10}
\begin{tabular}{rcrcr}
$k$ & \phantom{ABC} & $P(X_k < 0)$ & \phantom{ABC} & 
$P(X_k < 0\, | \,X_1 < 0)$ \\
\hline
1 & &      1/2 = 50.0\% & &       1 = 100.0\% \\
2 & &      1/4 = 25.0\% & &     1/2 = 50.0\% \\
3 & &      1/2 = 50.0\% & &     3/4 = 75.0\% \\
4 & &     5/16 = 31.2\% & &     1/2 = 50.0\% \\
5 & &      1/2 = 50.0\% & &   11/16 = 68.8\% \\
6 & &    11/32 = 34.4\% & &     1/2 = 50.0\% \\
7 & &      1/2 = 50.0\% & &   21/32 = 65.6\% \\
8 & &   93/256 = 36.3\% & &     1/2 = 50.0\% \\
9 & &      1/2 = 50.0\% & & 163/256 = 63.7\% \\
10 & & 193/512 = 37.7\% & &     1/2 = 50.0\% \\
\hline
\end{tabular}
\end{center}
\label{probtable}
\end{table}


We provide in Table \ref{probtable} the values of $P(X_k < 0)$, for $k \le 10$.  In all ten cases, the corresponding probability of Cash-Out is 0, a result that surely greatly relieves the CPDO, but we shall expand on that point later -- to the CPDO's added discomfiture.  

Extending Table \ref{probtable}, we now present a closed-form expression for $P(X_k < 0)$ for moderate values of $k$; this result will explain the regular appearance of the value 50\% in the table and show that it prevails for a moderately large number of tosses.  

\bigskip

\begin {theorem}
\label{xkprobs}
For $m=1,2,\ldots,199$,
\begin{equation}
\label{xevenprobs}
P(X_{2m} < 0) = \frac12 - \frac{1}{2^{2m+1}}\binom{2m}{m},
\end{equation}
and, for $m = 0,1,2,\ldots,99$, 
\begin{equation}
\label{xoddprobs}
P(X_{2m+1} < 0) = \frac12.
\end{equation}
\end {theorem}


A proof of this result is provided in Section \ref{proofs}.  

Another disconcerting property of the CPDO is its behavior conditional on having a loss on the first, or on an early toss.  Consider, say, $P(X_5 < 0 \, | \, X_1 < 0)$, the conditional probability that the CPDO will have a loss on the fifth toss given that it has a loss on the first toss.  By applying (\ref{xkexplicit}) and the independence of $C_1$ from $C_2,\ldots,C_5$, we obtain 
\begin{eqnarray}
\label{x5givenx1}
P(X_5 < 0 \, | \, X_1 < 0) & = & 
P\Big(\prod_{j=1}^5 (1-\delta C_j) > 1 \, | \, C_1 = -1\Big) 
\nonumber \\
& = & P\Big((1+\delta)\prod_{j=2}^5 (1-\delta C_j) > 1\Big).
\end{eqnarray}
Noting that $C_2, C_3, C_4, C_5$ are uniformly and independently distributed over $\{-1, +1\}$, and calculating the number of times that the inequality in (\ref{x5givenx1}) is satisfied, we obtain $P(X_5 < 0 \, | \, X_1 < 0) = 11/16$, or 68.8\%.  In like manner, we complete the last column of Table \ref{probtable}, and we also extend those calculations by deriving explicit formulas for the conditional probabilities $P(X_k < 0 \, | \, X_1 < 0)$ for higher values of $k$.  


\begin {theorem}
\label{xkconditional}
For $m = 1,2,\ldots,199$, 
\begin{equation}
\label{xkconditionaleven}
P(X_{2m} < 0 \, | \,X_1 < 0) = \frac12,
\end{equation}
and, for $m = 1,2,\ldots,99$, 
\begin{equation}
\label{xkconditionalodd}
P(X_{2m+1} < 0 \, | \, X_1 < 0) = \frac12 
+ \frac{1}{2^{2m+1}} \binom{2m}{m}.
\end{equation}
\end {theorem}

\bigskip

It is also the case that a loss at the second, or any early toss, leads to results similar to Theorem \ref{xkconditional}.  For instance, we shall derive explicit formulas for the conditional probabilities, $P(X_k < 0 \, | \, X_2 < 0)$, for higher values of $k$.

\bigskip

\begin {theorem}
\label{xkconditional2}
For $m = 1,2,\ldots,199$, 
\begin{equation}
\label{xkconditional2even}
P(X_{2m} < 0 \, | \, X_2 < 0) = 
\frac12 + \frac{1}{2^{2m-1}}\binom{2m-2}{m-1};
\end{equation}
and for $m = 1,2,\ldots,99$, 
\begin{equation}
\label{xkconditional2odd}
P(X_{2m+1} < 0 \, | \, X_2 < 0) = \frac12.
\end{equation}
\end {theorem}

\bigskip


Thus, the probabilities of net capital losses are substantially large for early tosses, as are the probabilities of future losses conditional on early losses, and these probabilities remain large for a considerable number of tosses.  

A related problem is the evaluation of $P(X_{k+1} < 0 \, | \, X_k < 0)$, the conditional probability that the CPDO will have a net capital loss at the $(k+1)$th toss, given that it has a loss at the $k$th toss.  Here, the results are startling and are nicely illustrated by the example of $P(X_3 < 0 \, | \, X_2 < 0)$:  Because $X_2 < 0$ if and only if $C_1 = C_2 = -1$, i.e., the first two tosses resulted in tails, then 
\begin{eqnarray*}
P(X_3 < 0 \, | \, X_2 < 0) & = & 
P((1-\delta C_1)(1-\delta C_2)(1-\delta C_3) > 1 | C_1=-1, C_2 = -1) \\
& = & P((1+\delta)^2 \, (1-\delta C_3) > 1) \\
& = & 1,
\end{eqnarray*}
because, with $\delta = 0.01$, 
$(1+\delta)^2 \, (1-\delta C_3) > 1$ for $C_3 = \pm 1$.  
It is interesting that the phenomenon $P(X_3 < 0 \, | \, X_2 < 0) = 1$ remains valid for all $\delta$ such that $(2+\delta)/(1+\delta)^2 > 1$, i.e., $0 < \delta < (\sqrt{5}-1)/2$, the reciprocal of the famous Golden Ratio.  

Generalizing the above example, we shall establish the following result that demonstrates astoundingly the lengthy effect of an early loss on the CPDO's net capital levels.  By this result, if the CPDO shows a net loss on an early toss then, for nearly two hundred additional tosses, the probability of a successive loss is 1 or close to 1.  

\bigskip

\begin {theorem}
\label{successive}
For $m = 1,2,3,\ldots,99$, 
\begin{equation}
\label{successiveeven}
P(X_{2m+1} < 0 \, | \, X_{2m} < 0) = 1,
\end{equation}
and for $m = 0,1,2,\ldots,98$, 
\begin{equation}
\label{successiveodd}
P(X_{2m+2} < 0 \, | \, X_{2m+1} < 0) 
= 1 - \frac{1}{2^{2m+1}}\binom{2m+1}{m+1}.
\end{equation}
\end {theorem}

\bigskip

We have studied, so far, the dependence between $X_{k+l}$ and $X_k$ for small values of $l$ only.  In the case of general $l$, we have the following result.  

\bigskip

\begin {theorem}
\label{xkplusrxl}
For $k,l=1,2,3,\ldots$, 
\begin{equation}
\label{memorylosses}
P(X_{k+l} < 0 | X_k < 0) \ge P(X_l < 0)
\end{equation}
and 
\begin{equation}
\label{memorycashout}
P(X_{k+l} \le -\gamma | X_k < 0) \ge P(X_l \le -\gamma).
\end{equation}
More generally, if $c_1 \in \R$, $c_2 > -1$, and 
$P(X_k < -c_2) > 0$ then 
\begin{equation}
\label{memorycashoutext}
P(X_{k+l} \le -c_1 | X_k < -c_2) \ge 
P\big(X_l < \frac{c_2-c_1}{1+c_2}\big).
\end{equation}
\end {theorem}

\bigskip

Here again, the consequences for a CPDO are ominous.  The inequality (\ref{memorylosses}) means that given that the CPDO has a loss at the $k$th toss, the probability that it has a loss at $l$th additional tosses can never be smaller than the (unconditional) probability of a loss at the $l$th toss.  Hence, the CPDO is punished in perpetuity following an early loss.  Similar remarks apply to the inequality (\ref{memorycashout}) which asserts that, conditional on a loss, the probability of Cash-Out at any subsequent stage is not smaller than the corresponding unconditional probability.

\subsection{Cash-Out: The case of the impatient banker}
\label{cashout}

\medskip

\hfill{\small \sl ``That which is painful, instructs.''}

\hfill{\small -- Benjamin Franklin}

\bigskip

Because not all bankers are infinitely patient, we now consider the case in which the CPDO receives from the banker a prompt Cash-Out call if its net assets falls to or below $-\gamma$.  

As we have shown in Section \ref{zeno}, it is impossible for the Cash-In event to occur in a finite number of tosses.  Consequently, we modify the Cash-In rule throughout this section of the paper to mean that the CPDO attains a net capital level of at least $1-\gamma$.  Noting that the initial net capital is $0$, if the CPDO attains modified Cash-In with $X_k \ge 1-\gamma$ then the net return on personal capital will be at least $(1-\gamma)/\gamma$; with $\gamma = 0.1$, this will result in a net return of at least $1,000$\%, an amount which is likely to satisfy any realistic CPDO noteholder.  


We shall need to determine the probability that the game continues forever, i.e., that the CPDO avoids both Cash-Out and Cash-In indefinitely.  Thus, we define 
\begin{equation}
\label{nocashinorout}
\alpha_k := P(-\gamma < X_1 < 1-\gamma,\ldots,-\gamma < X_k < 1-\gamma),
\end{equation}
the probability that the game lasts for at least $k$ tosses, i.e., the CPDO will avoid Cash-Out and Cash-In over the first $k$ tosses.  Then, we shall prove the following result.  

\bigskip

\begin{theorem}
\label{terminator}
Almost surely, the CPDO game terminates in a finite number of tosses.  That is, 
$$
\lim_{k \to \infty} \alpha_k = 0.
$$
\end{theorem}

\bigskip

Finally, we obtain an upper bound for the probability that the CPDO will attained Cash-In.  The result is likely to be painful, and hence instructive, for real-world CPDOs and so we present its proof immediately.  

\bigskip

\begin {theorem} 
\label{cashoutprob}
The probability that the CPDO attains Cash-In is at most 
$$
\gamma+\delta+\gamma\delta = 11.1\%.
$$
\end {theorem}

\noindent
{\it Proof}.  By Theorem \ref{terminator}, the game terminates in a finite number of tosses, almost surely.  Then, we define $P_w$ to be the probability that the CPDO wins, i.e., attains Cash-In.  Conditional on the event that the CPDO wins, let $E_w$ be the CPDO's expected net capital at that time; and, similarly, conditional on the event that the CPDO loses, i.e., attains Cash-Out, let $E_l$ be the CPDO's expected net capital at that time.  
  
The expected net capital for the CPDO after every toss is equal to the expected net capital before the toss.  The bankroll of the CPDO being bounded at all times, the expected net capital at the beginning and at the end of the game must be the same.  Therefore, by the Law of Total Probability, 
$$
0 = P_w E_w + (1-P_w) E_l.
$$
Solving this equation for $P_w$, we obtain 
\begin{equation}
\label{pwin}
P_w = -\frac{E_l}{E_w  - E_l}.
\end{equation}
If the CPDO loses, his last wager could not have exceeded $(1+\gamma)\delta$; thus, 
$$
-\gamma \ge E_l \ge  -\gamma - (1+\gamma)\delta = 
-(\gamma + \delta + \gamma\delta),
$$
equivalently, 
$$
\gamma \le -E_l \le \gamma + \delta + \gamma\delta.
$$
On the other hand, it is obvious that $E_w \ge 1-\gamma$, hence $E_w - E_l \ge 1-\gamma + \gamma = 1$.  Combining these bounds on $E_l$ and $E_w-E_l$ with equation (\ref{pwin}) gives 
$$
P_w = -\frac{E_l}{E_w  - E_l} 
\le \gamma + \delta + \gamma\delta = 0.111.
\qquad\Box
$$

\medskip

Simply put, even with the less stringent modified Cash-In rule, the probability that a CPDO will Cash-Out is at least 88.9\%.  We remark that the bound of 88.9\% likely is very close to the exact value of that probability; indeed, the results in Section \ref{simulations} of 1,000 CPDOs, each tossing their coin 50,000 times, starting with total capital of \$1,000, and subject to the modified Cash-In rule, indicates that the proportion of CPDOs that Cash-Out is close to 89\%.

\section{Conclusions}
\label{conclusions}
\setcounter{equation}{0}

{\parskip=-1.5pt
{\sl 
\hfill{\small The tale is far older than the Buttonwood Tree -- and new as IPO veneer --}

\hfill{\small For each trader knows ere his bonus grows he is master of Greed and Fear;}

\hfill{\small And each trader hears as midnight nears, and his bedsheets tug and chafe,}

\hfill{\small The Devil drum on the darkened pane: ``You've bought it, but is it Safe?''}

\smallskip
}
}

\hfill{\small -- with apologies to Rudyard Kipling}

\bigskip
\bigskip

Real-world CPDOs, invented at a time when the U.S. economy was near a zenith, were sold in August, 2006 as the economy weakened and began its descent into recession.  As credit default indices and ``bespoke'' securities weakened, Lady Luck thus handed CPDO buyers a string of tails from the start.  In an attempt to recover their losses, CPDO managers may have accelerated their dynamic hedging, tossing their coins faster, which gave them more tails.  It is even possible that the process of dynamic hedging {\it via} rapid and forced asset sales may have increased the probability of subsequent tails.  

The highly leveraged nature of CPDO notes, with notional amounts as high as fifteen times their nominal amounts, caused more grief.  At 15X leverage, CPDOs have $\gamma = 0.0625$ units of personal capital and $1 - \gamma = 0.9375$ units of borrowed capital; and their bankers, friend or foe, are understandably nervous about the tiny margin for allowable losses.  We also suspect that CPDO managers, on registering net capital losses at early tosses, increased $\delta$, the betting proportion, in a futile attempt to return to profits.  By (\ref{variance}), an increase in $\delta$ simply increases fluctuations in $X_k$ and, at a time when credit-default indices were falling, would have accelerated the CPDOs' losses.  

And so, down went the CPDOs.  The obituaries provided in \cite{capon,jones4,khasawneh} are especially instructful.  

In retrospect, we suspect that the problems encountered by Moody's were caused less by software errors and more by the flawed nature of the CPDOs' basic betting strategy.  And yet, there are other prominent factors that, in our view, contributed greatly to the CPDO debacle.  We describe four of them as follows: 

\smallskip

{\it First}, the assessments of CPDOs provided by the Bank of England \cite{boe}, the Bank for International Settlements \cite{bis}, and the European Central Bank \cite{ecb} seemed to have come too late to be of help to early purchasers of CPDO notes.  (Moreover, we can sympathize with a purchaser who interpreted as laudatory the statement \cite[p. 5]{boe}, ``Major innovations in securitisation can also be observed in the field of synthetic processes.  To be mentioned are, for example, Constant Proportion Debt Obligations (CPDOs), products that have received particular interest as they are highly rated and at the same time promise high interest.'')

And yet, CPDO purchasers would have benefited from a close reading of the deeper analyses provided in \cite[p. 199--200]{boe} where CPDOs were described, ominously, as ``apparently a `free lunch' for investors''; in \cite[p. 87]{ecb}; and especially \cite[Appendix D, pp. 60--72]{bis} and \cite{ubscpdoflaws}, where the warnings were severe.  

Indeed, sales of CPDOs could have been curtailed had the comprehensive assessment of CPDO risks in \cite[{\it loc. cit.}]{bis} been available in 2006.  These risks, including management fees and mark-to-market accounting, surely would have given potential buyers pause.  The Bank for International Settlements \cite[p. 62]{bis} noted that an ``obvious flaw in the first-generation CPDO design is its vulnerability to a legal form of front-running'', and opined that the CPDO market in 2007-2008 did not then ``appear to be large enough for front-running to be a problem.''  And yet, as any quant would agree, the efficient-market hypothesis would be violated were {\it any} legal form of front-running to remain unexploited by market participants.  We think that wide knowledge of the risk of front-running could have reduced the number of potential CPDO buyers.

\smallskip

{\it Second}, we did not incorporate into our analysis of CPDOs risks such as capital gains taxes, commissions on sales and purchases, management fees, mark-to-market accounting, potential front-running, or a coin that is slightly biased towards tails.  Had we done so, the outcome would have been considerably worse than those illustrated by our calculations.  
We encourage our readers to calculate the various probabilities in Section \ref{probability} for the case in which $p = 18/38$, the probability of success at the roulette-wheel game of red-and-black.  It is well-known \cite[p. 107]{billingsley} that, with timid or bold play, a casino gambler's probability of success decreases as $p$ changes from $1/2$ to $18/38$, and readers who perform similar calculations for the CPDO coin-tossing model will find that the probabilities of Cash-In also decrease.  

\smallskip

{\it Third}, as we explained in the Introduction, the SPVs used the proceeds from the sales of their CPDO notes to make a highly leveraged sale of an index of credit default swaps.  As it turns out, such an action in late 2006 was most unwise and reckless.  To explain, we first provide a concise, down-to-earth explanation of the nature of credit default swaps, taken from Herzog \cite[p. 50]{herzog}:

\begin{quote}
``Credit default swaps are essentially unregulated insurance policies covering the losses on securities in case a triggering event occurs.  Such an event could be a fire, a plane crash, or a mortgage foreclosure.  Financial institutions buy credit default swaps to protect themselves against the adverse effects of such events. In this sense default swaps are similar to fire insurance in the sense that a homeowner buys fire insurance to protect his investment in case his house burns down. However, unlike fire insurance, credit default swaps can also be used as a purely speculative `investment.'  In this case, credit default swaps are like buying insurance against the risk that my neighbor's house burns down.  Whereas with my own house, I have what insurance professionals call an `insurable interest,' with my neighbor's house I do not. The situation with credit default swaps is similar to bookies trading bets, with banks and hedge funds gambling on whether an investment (say, a collection of subprime mortgages bundled into a security) will succeed or fail.''
\end{quote}

Thus, in late 2006, the buyers of CPDO notes were undertaking the risks that go with being a highly leveraged, i.e., highly undercapitalized, insurance company.  Moreover, in purchasing an index of credit default swaps in late 2006, the CPDO note holders were insuring not simply one house in one city but {\it entire and adjoining} neighborhoods in the same city.  With subprime mortgage lenders making loans frenziedly, huge numbers of homeowners had become accustomed to playing with (financial) fireworks in their living rooms in late 2006.  Consequently, housing fires were not unlikely to occur.  Given that a fire in one house was guaranteed to ignite fires in adjacent houses, thence throughout the neighborhood, and eventually across the entire city, a CPDO note holder, in his role as an unwise insurance company or as a very na\"ive bookie, was doomed to bankruptcy.

\smallskip

{\it Fourth}, the quants created in the case of a CPDO a structure that violated the efficient-market hypothesis, an axiom of the field of ``financial engineering.''  It surely is not a good omen when researchers violate their own axioms.  Moreover, it strengthens the well-known complaints that quants, under pressure from investment bankers to devise structured products to be sold to the public, are incapable of maintaining even quasi-scientific standards; and that the executives and boards of directors of ratings agencies are not competent to assess the quants' activities.

In light of the results that we have derived, we fear that the on-going worldwide financial crisis will be prolonged by other arcane financial derivatives with structures as bizarre as CPDOs.  Here, we refer to credit derivative product companies, some of which carry leverage of up to 80-to-1; CDOs-squared and -cubed; and exotic synthetic CDOs.  Frankly, it is difficult to escape the conclusion \cite{capon} that some of those instruments are little more than ``dead men walking.''  We expect that second-generation CPDOs \cite{ubscpdoflaws} also will likely cause grief if they employ a betting strategy similar to that of their first-generation counterparts, or depend crucially on an assumption of mean-reversion, or in any way represent a ``free lunch.''  With negative outcomes from these instruments likely to impact the financial markets broadly, we expect that the markets' current distrust of ratings and regulatory agencies is likely to prevail for years to come.  

The ``don't-blame-the-quants'' opinion of \cite{shreve} notwithstanding, we believe that the quants indeed should shoulder the blame for the CPDO debacle and its unsettling effects on the financial markets.  Insofar as the losses on CPDOs are concerned, it would appear that investors were beguiled by the quants and their sirenical financial devices in much the same way that boaters on the Rhine were mesmerized by Lorelei\,\footnote{\href{http://ingeb.org/Lieder/ichweiss.html}{The mythical Germanic siren} who, sitting on a high cliff while combing her beautiful hair and \href{http://www.loreley-info.com/eng/loreley/loreley-song.mp3}{singing sensually}, tempted boaters on the River Rhine to their deaths.  \href{http://www.websters-online-dictionary.org/Lo/Lorelei.html}{Webster's Online Dictionary} notes that ``\,`Lorelei' is a name that signifies or is derived from: `an ambush cliff'.''} and her captivating voice.



\section{Proofs of theorems}
\label{proofs}
\setcounter{equation}{0}

\noindent
{\it Proof of Theorem \ref{xkformula}}.  
By the recurrence formula (\ref{xkrecurrence}) we see that, for 
$k = 1,2,3,\ldots$, 
$$
1-X_k = 1 - X_{k-1} - \delta(1-X_{k-1})C_k 
= (1 - X_{k-1})(1 - \delta C_k),
$$
On applying this formula with $k$ replaced by $k-1,k-2,\ldots,1$, we obtain 
\begin{eqnarray*}
1-X_k & = & (1 - X_{k-1})(1 - \delta C_k) \\
& = & (1 - X_{k-2})(1 - \delta C_{k-1})(1 - \delta C_k) \\
& \vdots & \\
& = & (1-X_0) (1-\delta C_1) (1-\delta C_2) \cdots 
(1-\delta C_{k-1})(1 - \delta C_k).
\end{eqnarray*}
Substituting $X_0 = 0$ and solving for $X_k$, we obtain (\ref{xkexplicit}).
$\qed$

\bigskip
\bigskip

\noindent
{\it Proof of Theorem \ref{expectxk}}.  
Noting that $C_1,\ldots,C_k$ are mutually independent and, by (\ref{cj}), that $E(C_j) = 0$ for all $j$, it follows from (\ref{xkexplicit}) that 
$$
E(X_k) = 1 - \prod_{j=1}^k E(1-\delta C_j) 
= 1 - \prod_{j=1}^k \big(1-\delta E(C_j)\big) = 0.
$$  
Consequently, 
\begin{eqnarray*}
\Var(X_k) & = & \Var\prod_{j=1}^k (1-\delta C_j) \\
& = & E\prod_{j=1}^k (1-\delta C_j)^2 
- \big(E\prod_{j=1}^k (1-\delta C_j)\big)^2 \\
& = & \big(\prod_{j=1}^k E(1-\delta C_j)^2\big) - 1.
\end{eqnarray*}
Since $C_j^2 \equiv 1$ and $E(C_j) = 0$, we have 
$E(1 - \delta C_j)^2 = E(1- 2 \delta C_j + \delta^2 C_j^2) 
= 1+\delta^2$, and then (\ref{variance}) follows immediately.

Finally, (\ref{variance}) implies that for large $k$, 
$\Var(X_k) \sim \exp(k\ln (1+\delta^2))$, 
so $\Var(X_k)$ increases exponentially with $k$.
$\qed$

\bigskip
\bigskip

\noindent
{\it Proof of Theorem \ref{losses}}.  
By (\ref{xkexplicit}), the inequality $X_k \ge t$ is equivalent to 
\begin{equation}
\label{xkyk}
\prod_{j=1}^k (1-\delta C_j) \le 1-t.
\end{equation}
Let $W_j = \ln(1-\delta C_j)$, where $j=1,\ldots,k$; by taking natural logarithms in the inequality (\ref{xkyk}), we see that (\ref{xkyk}) is equivalent to $\sum_{j=1}^k W_j \le \ln(1-t)$.  Therefore, 
\begin{equation}
\label{xkyk2}
P(X_k \ge t) = P\Big(\sum_{j=1}^k W_j \le \ln(1-t)\Big).
\end{equation}
Since $C_1,\ldots,C_k$ are independent, identically distributed random variables, then so are $W_1,\ldots,W_k$.  Further, it is simple to show that the mean and variance of each $W_j$ are 
$$
\mu = \tfrac12\ln(1-\delta^2)
\quad 
\hbox{and}
\quad
\sigma^2 = \tfrac14\Big(\ln\frac{1+\delta}{1-\delta}\Big)^2,
$$
respectively, and $\mu < 0$ because $\delta < 1$.  

Let $\Phi(\cdot)$ 
denote the cumulative distribution function of the standard normal distribution.  On applying the Central Limit Theorem \cite[p. 434]{ross} to $\sum_{j=1}^k W_j$, we obtain 
\begin{equation}
\label{clt}
\lim_{k \to \infty} P\Big(\sum_{j=1}^k W_j \le \ln(1-t)\Big) = 
\lim_{k \to \infty} 
\Phi\Big(\frac{\ln(1-t)-k\mu}{\sqrt{k\sigma^2}}\Big) = 1.
\end{equation}

At this stage we can verify the remark, made below the statement of Theorem \ref{losses}, regarding the number of tosses necessary for the CPDO to attain $X_k \ge 1-\gamma$ with probability 95\%.  Indeed, on applying the approximation, 
$$
P(X_k \ge t) \simeq \Phi\Big(\frac{\ln(1-t)-k\mu}{\sqrt{k\sigma^2}}\Big),
$$
which holds by (\ref{xkyk2}) and (\ref{clt}), equating the right-hand side of this approximation to 0.95 and solving for large $k$, we obtain $k \simeq 108,218.13$.  


Next, we note that $\{C_1 = -1,\ldots,C_k = -1\}$ implies $\{X_k < 0\}$.  Therefore,  
$$
P(X_k < 0) \ge P(C_1 = -1,\ldots,C_k = -1) = 2^{-k} > 0,
$$
so any conditional probability that is conditioned on the event $\{X_k < 0\}$ is well-defined.  

By definition, 
\begin{equation}
\label{kconditional}
P(X_{k+1} < 0 \, | \, X_k < 0) = 
\frac{P(X_{k+1} < 0, X_k < 0)}{P(X_k < 0)}.
\end{equation}
By the Law of Total Probability \cite[p. 72]{ross}, 
\begin{equation}
\label{totalprob}
\begin{aligned}
P\big(X_{k+1} < 0, X_k < 0\big) = \ & 
P\big(X_{k+1} < 0, X_k < 0 \, | \, C_{k+1} = -1\big) P(C_{k+1} = -1) \\
& + \, P\big(X_{k+1} < 0, X_k < 0 \, | \, C_{k+1} = +1\big) 
P(C_{k+1} = +1).
\end{aligned}
\end{equation}
By (\ref{xkrecurrence}), 
\begin{eqnarray*}
P\big(X_{k+1} < 0, X_k < 0 \, | \, C_{k+1} = -1\big) & = & 
P\big(X_k - \delta(1-X_k) < 0, X_k < 0\big) \\
& = & P(X_k < 0),
\end{eqnarray*}
and also, 
\begin{eqnarray*}
P\big(X_{k+1} < 0, X_k < 0 \, | \, C_{k+1} = +1\big) & = & 
P\big(X_k + \delta(1-X_k) < 0, X_k < 0\big) \\
& = & P\big(X_k < -\frac{\delta}{1-\delta}\big).
\end{eqnarray*}
Therefore, by equations (\ref{kconditional}) and (\ref{totalprob}), 
\begin{eqnarray}
\label{xktoxkplus}
P(X_{k+1} < 0 \, | \, X_k < 0) & = & 
\frac{P\big(X_k < 0\big) + 
P\big(X_k < -\frac{\delta}{1-\delta}\big)}{2 P(X_k < 0)} \nonumber \\
& = & \frac12 
\Big[1 + \frac{P\big(X_k < -\frac{\delta}{1-\delta}\big)}{P(X_k < 0)}\Big].
\end{eqnarray}
For large values of $k$, it follows from (\ref{clt}) that 
\begin{equation}
\label{cltapprox}
P\big(X_k < -\frac{\delta}{1-\delta}\big) \simeq 
\Phi\Big(\frac{\ln(1-\delta) + k\mu}{\sqrt{k\sigma^2}}\Big)
\end{equation}
and 
$$
P(X_k < 0) \simeq \Phi\Big(\frac{k\mu}{\sqrt{k\sigma^2}}\Big).
$$
Therefore, 
$$
\lim_{k \to \infty} 
\frac{P\big(X_k < -\frac{\delta}{1-\delta}\big)}{P(X_k < 0)} = 
\lim_{k \to \infty} 
\frac{\Phi\Big(\big(\ln(1-\delta) + k\mu\big)/\sqrt{k\sigma^2}\Big)}
{\Phi\big(k\mu/\sqrt{k\sigma^2}\big)} \\
= 1,
$$
where the second equality is obtained by an application of L'Hospital's Rule.  Consequently, it follows from (\ref{xktoxkplus}) that 
$\lim_{k \to \infty} P(X_{k+1} < 0 \, | \, X_k < 0) = 1$.  

The proof that 
$\lim_{k \to \infty} P(X_{k+1} \le -\gamma \, | \, X_k < 0) = 1$ 
proceeds by a similar argument.  By the same approach that led to (\ref{xktoxkplus}), we obtain 
$$
P(X_{k+1} \le -\gamma \, | \, X_k < 0) = 
\frac{P\big(X_k \le \frac{-\gamma+\delta}{1+\delta}\big) + 
P\big(X_k < \frac{-\gamma-\delta}{1-\delta}\big)}{2 P(X_k < 0)}.
$$
On approximating each term on the right-hand side as in (\ref{cltapprox}), and then applying L'Hospital's rule, we obtain 
$\lim_{k \to \infty} P(X_{k+1} \le -\gamma \, | \, X_k < 0) = 1$.  
$\qed$

\bigskip
\bigskip

\noindent
{\it Proof of Theorem \ref{martingale}}.  
In calculating $X_k$ from earlier outcomes $\{X_1,\ldots,X_{k-1}\}$, it follows from equation (\ref{xkrecurrence}) that only $X_{k-1}$ and $C_k$ are germane.  Because $C_k$ is independent of $X_{k-1}$ and $E(C_k) = 0$, we obtain 
\begin{eqnarray*}
E(X_k\, | \,X_1,\ldots,X_{k-1}) & = & E(X_k \, | \, X_{k-1}) \\
& = & X_{k-1} + \delta(1-X_{k-1})E(C_k) = X_{k-1}.
\qquad\Box
\end{eqnarray*}

\bigskip
\bigskip

\noindent
{\it Proof of Theorem \ref{xkprobs}}.  
By (\ref{xkexplicit}), 
$$
P(X_{2m} < 0) = P\big(\prod_{j=1}^{2m} (1-\delta C_j) > 1\big).
$$
Suppose that $i$ of the random variables $C_1,\ldots,C_{2m}$ are equal to $-1$ and all remaining $C_j$ are equal to $+1$.  Then we need to determine all values of $i$ such that the product 
$\prod_{j=1}^{2m} (1-\delta C_j) \equiv 
(1-\delta)^{2m-i}(1+\delta)^i > 1$.  
To that end, it is simple to verify that the sequence 
$$
a_i = (1-\delta)^{2m-i}(1+\delta)^i,
$$
where $i=0,\ldots,2m$, is strictly increasing as $i$ increases; in fact, 
$$
\frac{a_{i+1}}{a_i} = \frac{1+\delta}{1-\delta} > 1,
$$
for all $i=0,\ldots,2m-1$.  Note also that 
$a_m = (1-\delta)^m (1+\delta)^m = (1-\delta^2)^m < 1$, 
so $a_i < 1$ for all $i=0,\ldots,m$, and therefore the first value of $i$ for which it is possible to have $a_i > 1$ is $i = m+1$.  By solving the inequality $a_{m+1} > 1$, i.e., 
$(1-\delta)^{m-1} (1+\delta)^{m+1} > 1$, we obtain $m \le 199$.  Consequently, for such $m$, 
$$
(1-\delta)^{2m-i}(1+\delta)^i 
\begin{cases}
< 1, & i=0,\ldots,m, \\
> 1, & i=m+1,\ldots,2m.
\end{cases}
$$
Next, for each $i \ge m+1$, we count the number of cases in which $i$ of $C_1,\ldots,C_{2m}$ are equal to $-1$ and the remaining $2m-i$ variables $C_j$ are equal to $+1$.  This is a standard calculation \cite[p. 6]{ross} in an undergraduate course on probability theory; the answer is the binomial coefficient, 
$$
\binom{2m}{i} = \frac{(2m)!}{(2m-i)! \, i!}.
$$
Because there are $2^{2m}$ equally likely possible choices for the sequence $C_1,\ldots,C_{2m}$, it follows that, for $m \le 199$, 
\begin{equation}
\label{xevens}
P(X_{2m} < 0) = \frac{1}{2^{2m}} 
\sum_{i=m+1}^{2m} \binom{2m}{i}.
\end{equation}
This sum is well-known; once we recognize that it is the sum of nearly the last half of all entries in the $2m$-th row of Pascal's triangle, we deduce that it equals 
$$
2^{2m-1} - \frac12 \binom{2m}{m}.
$$
On substituting this result into (\ref{xevens}), we obtain (\ref{xevenprobs}).

For the case in which $k$ is odd, say, $k = 2m+1$, the calculations are similar.  By applying the same arguments as before we deduce that, for all $m \le 99$, 
$$
P(X_{2m+1} < 0) = \frac{1}{2^{2m+1}} 
\sum_{i=m+1}^{2m+1} \binom{2m+1}{i};
$$
noting that this sum is precisely the sum of the last half of all numbers in the $(2m+1)$-th row of Pascal's triangle, we find that it equals $2^{2m+1}/2 = 2^{2m}$.  Therefore, for $m \le 99$, $P(X_{2m+1} < 0) = 2^{2m}/2^{2m+1} = 1/2$.  
$\qed$

\bigskip
\bigskip

\noindent
{\it Proof of Theorem \ref{xkconditional}}.  
Let $k = 2m$, where $m$ is a positive integer.  By the same argument as at (\ref{x5givenx1}), 
\begin{equation}
\label{x2mgivenx1}
P(X_{2m} < 0 \, | \, X_1 < 0) 
= P\Big((1+\delta)\prod_{j=2}^{2m} (1-\delta C_j) > 1\Big).
\end{equation}
To calculate this probability, we need to enumerate the number of cases in which the inequality on the right-hand side of (\ref{x2mgivenx1}) is satisfied as each of $C_2,\ldots,C_{2m}$ varies uniformly and independently over the set $\{-1, +1\}$.  

Suppose that $i$ of the variables $C_2,\ldots,C_{2m}$ are equal to $-1$ and all remaining $C_j$ are equal to $+1$; then we need to determine all values of $i$ for which the product $(1+\delta)\prod_{j=2}^{2m} (1-\delta C_j) \equiv (1-\delta)^{2m-1-i}(1+\delta)^{i+1} > 1$.  Now, it is simple to verify that the sequence 
$$
a_i = (1-\delta)^{2m-1-i}(1+\delta)^{i+1},
$$
where $i=0,\ldots,2m-1$, is strictly increasing as $i$ increases; in fact, 
$$
\frac{a_{i+1}}{a_i} = \frac{1+\delta}{1-\delta} > 1,
$$
for all $i=0,\ldots,2m-1$.  Also, note that 
$a_{m-1} = (1-\delta)^m (1+\delta)^m = (1-\delta^2)^m < 1$, 
so $a_i < 1$ for all $i=0,\ldots,m-1$, and therefore the first value of $i$ for which it is possible to have $a_i > 1$ is $i = m$.  By solving the inequality $a_m > 1$, i.e., 
$(1-\delta)^{m-1} (1+\delta)^{m+1} > 1$, we obtain $m \le 199$, in which case 
$$
(1-\delta)^{2m-1-i}(1+\delta)^{i+1} 
\begin{cases}
< 1, & i=0,\ldots,m-1, \\
> 1, & i=m,\ldots,2m-1.
\end{cases}
$$
Next, for each $i \ge m$, we count the number of cases in which $i$ of $C_2,\ldots,C_{2m}$ are equal to $-1$ and the remaining $2m-1-i$ variables $C_j$ are equal to $+1$.  Similar to the foregoing, the answer is the binomial coefficient, 
$$
\binom{2m-1}{i} = \frac{(2m-1)!}{(2m-1-i)! \, i!}.
$$  
Because there are $2^{2m-1}$ equally likely possible choices for the sequence $C_2,\ldots,C_{2m}$, it follows that, for $m \le 199$, 
$$
P(X_{2m} < 0 \, | \, X_1 < 0) = \frac{1}{2^{2m-1}} 
\sum_{i=m}^{2m-1} \binom{2m-1}{i}.
$$
This sum is well-known, being the sum of the last half of all entries in the $(2m-1)$th row of Pascal's triangle and, therefore, equals $2^{2m-1}/2 = 2^{2m-2}$.  Consequently, for $m \le 199$, 
$P(X_{2m} < 0\, | \,X_1 < 0) = 2^{2m-2}/2^{2m-1} = 1/2$.  

For the case in which $k$ is odd, say, $k = 2m+1$, the calculations are similar.  We deduce, first, that for all $m \le 99$, 
$$
P(X_{2m+1} < 0 \, | \, X_1 < 0) = \frac{1}{2^{2m}} 
\sum_{i=m}^{2m} \binom{2m}{i};
$$
second, by comparing this sum to the total of the last half of all numbers in the $2m$-th row of Pascal's triangle, we obtain 
\begin{equation}
\label{pascaleven}
\sum_{i=m}^{2m} \binom{2m}{i} = 2^{2m-1} + \frac12\binom{2m}{m};
\end{equation}
and this leads to the stated result.  
$\qed$

\bigskip
\bigskip

\noindent
{\it Proof of Theorem \ref{xkconditional2}}.  The proof of this result is similar to the proof of Theorem \ref{xkconditional}.  Nevertheless, we provide complete details in order that the paper be self-contained.

Let $k = 2m$, where $m$ is a positive integer, $m \ge 2$.  By the same argument as at (\ref{x5givenx1}), 
\begin{equation}
\label{x2mgivenx2}
P(X_{2m} < 0 \, | \, X_2 < 0) 
= P\Big((1+\delta)^2\prod_{j=3}^{2m} (1-\delta C_j) > 1\Big).
\end{equation}
To calculate this probability, we need to enumerate the number of cases in which the inequality on the right-hand side of (\ref{x2mgivenx2}) is satisfied as each of $C_3,\ldots,C_{2m}$ varies uniformly and independently over the set $\{-1, +1\}$.  

Suppose that $i$ of the variables $C_3,\ldots,C_{2m}$ are equal to $-1$ and all remaining $C_j$ are equal to $+1$; then we need to determine all values of $i$ for which the product 
$(1+\delta)^2\prod_{j=3}^{2m} (1-\delta C_j) \equiv 
(1-\delta)^{2m-2-i}(1+\delta)^{i+2} > 1$.  
Now, it is simple to verify that the sequence 
$$
a_i = (1-\delta)^{2m-2-i}(1+\delta)^{i+2},
$$
where $i=0,\ldots,2m-2$, is strictly increasing as $i$ increases; in fact, 
$$
\frac{a_{i+1}}{a_i} = \frac{1+\delta}{1-\delta} > 1,
$$
for all $i=0,\ldots,2m-2$.  Also, note that 
$a_{m-2} = (1-\delta)^m (1+\delta)^m = (1-\delta^2)^m < 1$, 
so $a_i < 1$ for all $i=0,\ldots,m-2$, and therefore the first value of $i$ for which it is possible to have $a_i > 1$ is $i = m-1$.  By solving the inequality $a_{m-1} > 1$, i.e., 
$(1-\delta)^{m-1} (1+\delta)^{m+1} > 1$, we obtain $m \le 199$.  For any such $m$, 
$$
(1-\delta)^{2m-2-i}(1+\delta)^{i+2} 
\begin{cases}
< 1, & i=0,\ldots,m-2, \\
> 1, & i=m-1,\ldots,2m-2.
\end{cases}
$$
Next, for each $i \ge m-1$, the number of cases in which $i$ of $C_3,\ldots,C_{2m}$ are equal to $-1$ and the remaining $2m-2-i$ variables $C_j$ all are equal to $+1$ is the binomial coefficient, 
$$
\binom{2m-2}{i} = \frac{(2m-2)!}{(2m-2-i)! \, i!}.
$$  
Because there are $2^{2m-2}$ equally likely possible choices for the sequence $C_3,\ldots,C_{2m}$, it follows that 
$$
P(X_{2m} < 0 \, | \, X_2 < 0) = \frac{1}{2^{2m-2}} 
\sum_{i=m-1}^{2m-2} \binom{2m-2}{i}.
$$
On evaluating this sum by means of (\ref{pascaleven}) and substituting its value into the previous equation, we obtain the desired result.  

For the case in which $k$ is odd, say, $k = 2m+1$ with $m \ge 1$, we deduce by a similar argument that, for $m \le 99$, 
$$
P(X_{2m+1} < 0 \, | \, X_2 < 0) = \frac{1}{2^{2m-1}} 
\sum_{i=m-1}^{2m-1} \binom{2m-1}{i} = \frac12.
\qquad\Box
$$

\bigskip
\bigskip

\noindent
{\it Proof of Theorem \ref{successive}}.  
By the Law of Total Probability, 
\begin{eqnarray*}
P(X_{k+1} < 0 \, | \, X_k < 0) & = &
P(X_{k+1} < 0 \, | \, X_k < 0, C_{k+1} = -1) P(C_{k+1} = -1) \\
& & \ + \ P(X_{k+1} < 0 \, | \, X_k < 0, C_{k+1} = +1) P(C_{k+1} = +1) \\
& = & \tfrac12[P(X_{k+1} < 0 \, | \, X_k < 0, C_{k+1} = -1) \\
& & \ + \ P(X_{k+1} < 0 \, | \, X_k < 0, C_{k+1} = +1)].
\end{eqnarray*}
It is clear that a net capital loss at the $k$th stage followed by tails at the next toss guarantees a net loss at the $(k+1)$th stage.  Therefore, 
$$
P(X_{k+1} < 0 \, | \, X_k < 0, C_{k+1} = -1) = 1,
$$ 
and so we have 
\begin{equation}
\label{hydra}
P(X_{k+1} < 0 \, | \, X_k < 0) = \frac12\Big(1 + 
P(X_{k+1} < 0 \, | \, X_k < 0, C_{k+1} = +1)\Big).
\end{equation}

Next, by the definition of conditional probability,  
$$
P(X_{k+1} < 0 \, | \, X_k < 0, C_{k+1} = +1) = 
\frac{P(X_{k+1} < 0, X_k < 0, C_{k+1} = +1)}
{P(X_k < 0, C_{k+1} = +1)},
$$
If $C_{k+1} = +1$ then, obviously, $X_{k+1} > X_k$; we see now that the events $\{X_{k+1} < 0, X_k < 0, C_{k+1} = +1\}$ and $\{X_{k+1} < 0, C_{k+1} = +1\}$ are equivalent, and hence have 
the same probability.  Therefore, 
\begin{equation}
\label{conditionalhydra}
P(X_{k+1} < 0 \, | \, X_k < 0, C_{k+1} = +1) = 
\frac{P(X_{k+1} < 0, C_{k+1} = +1)}{P(X_k < 0, C_{k+1} = +1)}.
\end{equation}
We proceed now to evaluate the numerator and denominator above.  Because $X_k$ and $C_{k+1}$ are mutually independent, it follows from Theorem \ref{xkprobs} that 
\begin{eqnarray*}
P(X_k < 0, C_{k+1} = +1) & = & \frac12 P(X_k < 0) \\
& = &  \begin{cases}
\frac12\Big(\frac12 - \frac{1}{2^{2m+1}}\binom{2m}{m}\Big), & 
k = 2m, \ 1 \le m \le 199, \\
\frac14, & k = 2m+1, \ 0 \le m \le 99.
\end{cases}
\end{eqnarray*}
As for the numerator in (\ref{conditionalhydra}), it follows by (\ref{xkexplicit}) and the mutual independence of the $C_j$ that 
\begin{eqnarray*}
P(X_{k+1} < 0, C_{k+1} = +1) 
& = & P\big((1-\delta)\prod_{j=1}^k (1-\delta C_j) > 1, C_{k+1} = +1\big) \\ 
& = & \tfrac12 
P\big((1-\delta)\prod_{j=1}^k (1-\delta C_j) > 1\big).
\end{eqnarray*}
We calculate this probability by proceeding as in Theorem \ref{xkconditional}, thereby obtaining 
$$
P(X_{k+1} < 0, C_{k+1} = +1) = 
\begin{cases}
\frac12\Big(\frac12 - \frac{1}{2^{2m+1}}\binom{2m}{m}\Big), 
& k = 2m, \ 1 \le m \le 99, \\
\frac12\Big(\frac12 - \frac{1}{2^{2m+1}}\binom{2m+1}{m+1}\Big), & k = 2m+1, \ 1 \le m \le 198.
\end{cases}
$$

For $k = 2m$ with $1 \le m \le 99$, the numerator and denominator of (\ref{conditionalhydra}) are equal; then, by  (\ref{hydra}) and (\ref{conditionalhydra}), 
$P(X_{k+1} < 0 \,|\, X_k < 0) = 1$.  

For $k = 2m+1$ where $1 \le m \le 98$, we use the above results to obtain 
$$
P(X_{2m+2} < 0 \,|\, X_{2m+1} < 0, C_{2m+2} = +1) = 
1 - \frac{1}{2^{2m}}\binom{2m+1}{m+1},
$$
and then it follows from (\ref{hydra}) that 
$$
P(X_{2m+2} < 0 \,|\, X_{2m+1} < 0) = 1 - \frac{1}{2^{2m+1}}\binom{2m+1}{m+1}.
\qquad\Box
$$

\bigskip
\bigskip

\noindent
{\it Proof of Theorem \ref{xkplusrxl}}.  
Because (\ref{memorylosses}) and (\ref{memorycashout}) are special cases of (\ref{memorycashoutext}), it suffices to prove the latter only.

If $\prod_{j=1}^k (1-\delta C_j) > 1+c_2$ and 
$\prod_{j=k+1}^{k+l} (1-\delta C_j) \ge (1+c_1)/(1+c_2)$ 
then, clearly, $\prod_{j=1}^{k+l} (1-\delta C_j) \ge 1+c_1$.  Therefore, 
\begin{align*}
P\Big(\prod_{j=1}^{k+l} (1-\delta C_j) \ge \ 1 & +c_1,
\prod_{j=1}^k (1-\delta C_j) > 1+c_2 \Big) \\ 
& \ge \ P\Big(\prod_{j=k+1}^{k+l} (1-\delta C_j) \ge \frac{1+c_1}{1+c_2},
\prod_{j=1}^k (1-\delta C_j) > 1+c_2\Big) \\
& = \ P\Big(\prod_{j=k+1}^{k+l} (1-\delta C_j) \ge \frac{1+c_1}{1+c_2}\Big) 
P\Big(\prod_{j=1}^k (1-\delta C_j) > 1+c_2\Big) \\
& = \ P\Big(\prod_{j=1}^l (1-\delta C_j) \ge \frac{1+c_1}{1+c_2}\Big) 
P\Big(\prod_{j=1}^k (1-\delta C_j) > 1+c_2\Big),
\end{align*}
where the last two equalities are due to $C_1,\ldots,C_{k+l}$ being independent, identically distributed random variables.  Applying (\ref{xkexplicit}) to express the above inequalities in terms of the $X_k$, we obtain 
\begin{eqnarray*}
P(X_{k+l} \le -c_1, X_k < -c_2) & \le & 
P\big(1-X_l \ge \frac{1+c_1}{1+c_2}\big) P(X_k < -c_2) \\
& = & P\big(X_l \le \frac{c_2-c_1}{1+c_2}\big) P(X_k < -c_2),
\end{eqnarray*}
and this result is equivalent to (\ref{memorycashoutext}).  
$\qed$

\bigskip

Finally, we consider Theorem \ref{terminator}.  To prove that result, we require a preliminary calculation.  

\bigskip

\begin{lemma}
\label{choosebound} 
If k is an even positive integer then, for $j=0,\ldots,k$, 
$$
{k \choose j} < 2^k \Big(\frac{2}{\pi k}\Big)^{1/2} 
\exp\Big(\frac{1}{12 k}\Big).
$$
\end{lemma}

\noindent
{\it Proof}.  By Stirling's inequalities \cite[p. 43]{ross}, we have 
$$
(2 \pi k)^{1/2} \, k^k \, \exp(-k) < k! < (2 \pi k)^{1/2} \, k^k 
\exp\Big(-k + \frac1{12 k}\Big).
$$
Also, it is well-known \cite[p. 156]{ross} that 
$$
\max_{j=0,1,\ldots,k}{k \choose j} = {k \choose k/2} = 
\frac{k!}{\left((k/2)!\right)^2}.
$$
Therefore, for $j=0,\ldots,k$, 
\begin{eqnarray*}
{k \choose j} 
& \leq & \frac{k!}{\left((k/2)!\right)^2} \\
& < & \frac{(2 \pi k)^{1/2} \, k^k \, \exp(-k + \frac1{12 k})}
         {\big(
           (2 \pi k/2)^{1/2} \, (k/2)^{k/2} \, \exp(-k/2)
         \big)^2} \\
&=& 2^k \Big(\frac{2}{\pi k}\Big)^{1/2} \exp\Big(\frac{1}{12 k}\Big).
\qquad\qquad\square
\end{eqnarray*}

\bigskip
\bigskip

\noindent
{\it Proof of Theorem \ref{terminator}}.  
Let 
$N_k$ be the number of heads obtained among the first $k$ tosses of the coin.  Then
\begin{eqnarray*}
X_k & = & 1- \prod_{i=1}^k (1-\delta C_i)\\
    & = & 1 - (1-\delta)^{N_k} (1+\delta)^{k-N_k}\\
    & = & 1 - (1+\delta)^k\, \Big(\frac{1-\delta}{1+\delta}\Big)^{N_k}.
\end{eqnarray*}
Therefore, $-\gamma < X_k < 1-\gamma$ if and only if 
$$
-\gamma < 1 - (1+\delta)^k \, \Big(\frac{1-\delta}{1+\delta}\Big)^{N_k} < 1-\gamma.
$$
Solving this inequality for $N_k$ yields
$$
\frac{k\ln(1+\delta) - \ln(1+\gamma)}{\ln(1+\delta) - \ln(1-\delta)} < N_k < 
\frac{k\ln(1+\delta) - \ln\gamma}{\ln(1+\delta) - \ln(1-\delta)},
$$
and the length of this interval range for $N_k$ is 
\begin{equation}
\label{bineq}
\frac{\ln(1+\gamma) - \ln\gamma}{\ln(1+\delta) - \ln(1-\delta)},
\end{equation}
a finite number that we denote by $\beta$.  Hence, 
\begin{equation}
\label{boundpxk}
P(-\gamma < X_k < 1-\gamma) \leq \beta \max_{j=0,\ldots,k} P(N_k=j).
\end{equation}
Because $N_k$ has a binomial distribution, $N_k \sim B(k,1/2)$, we obtain 
\begin{equation}
\label{pnbound}
P(N_k=j) = 2^{-k} \, {k\choose j} 
< \Big(\frac{2}{\pi k}\Big)^{1/2} \exp\Big(\frac{1}{12 k}\Big),
\end{equation}
where the inequality in (\ref{pnbound}) follows from Lemma \ref{choosebound} when $k$ is even.  From inequalities (\ref{boundpxk}) and (\ref{pnbound})
we conclude that
\begin{equation}
\label{endprob1}
P(-\gamma < X_k < 1-\gamma) < \beta \, \Big(\frac{2}{\pi k}\Big)^{1/2}
\exp\Big(\frac{1}{12 k}\Big)  
\end{equation}
when $k$ is even.  By (\ref{nocashinorout}), $\alpha_k$ is monotonic decreasing and 
$$
0 \le \alpha_k \leq P(-\gamma < X_k < 1-\gamma);
$$
therefore, by (\ref{endprob1}), $\alpha_k \to 0$ as $k \to \infty$.
$\qed$

\vskip 1truein

\section{Results of 1,000 simulated CPDOs}
\label{simulations}

In the following figures, we  present the graphical results of the simulations of 1,000 CPDOs, each tossing their coin 50,000 times, starting with total capital of \$1,000, and subject to the modified Cash-In rule.  As the simulations indicate, the proportion of CPDOs that Cash-Out is close to 89\%.

\clearpage

\begin{figure} 
  \includegraphics[width=\textwidth]{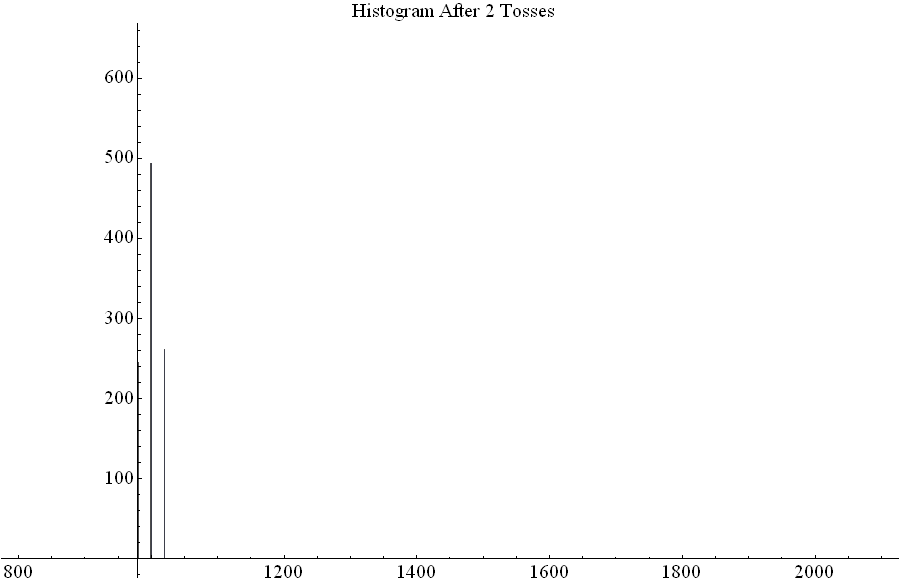}
\end{figure}

\clearpage

\begin{figure} 
  \includegraphics[width=\textwidth]{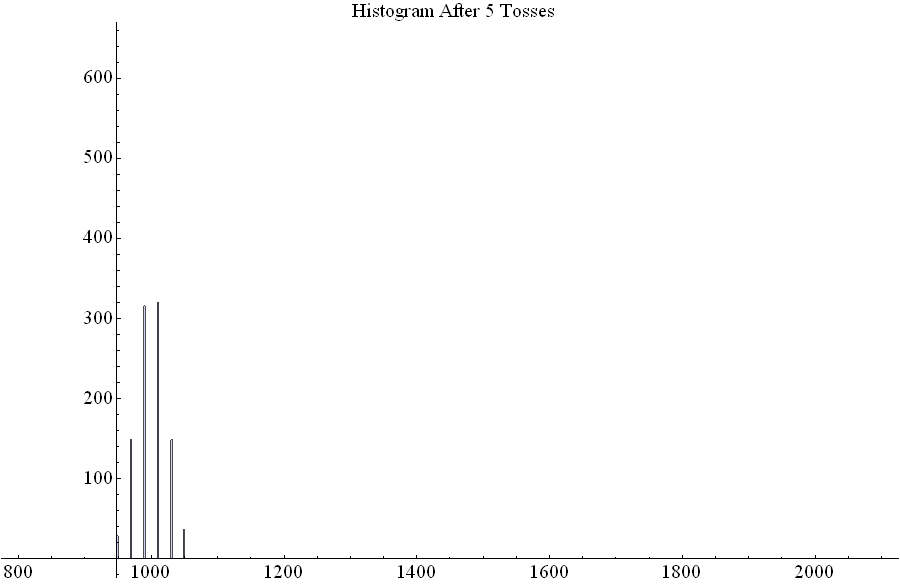}
\end{figure}

\clearpage

\begin{figure} 
  \includegraphics[width=\textwidth]{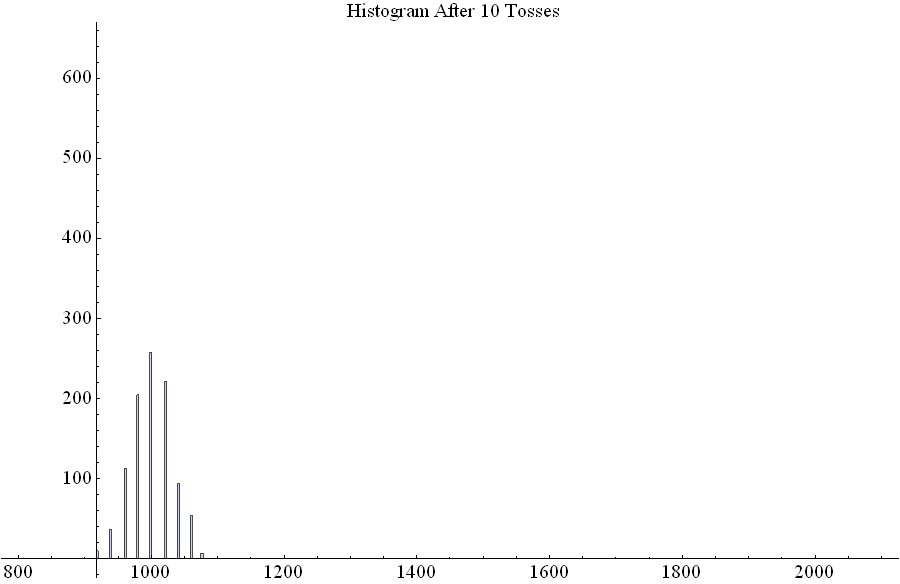}
\end{figure}

\clearpage

\begin{figure} 
  \includegraphics[width=\textwidth]{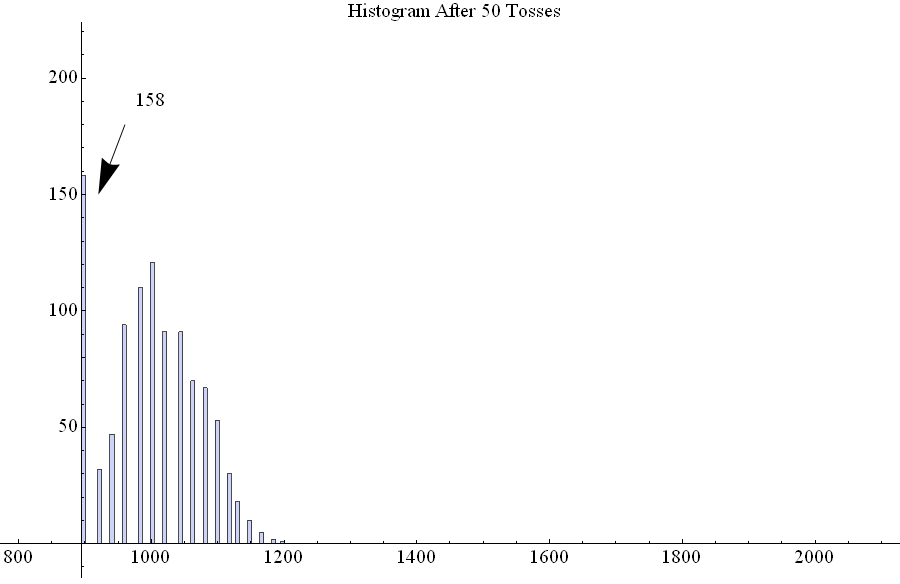}
\end{figure}

\clearpage

\begin{figure} 
  \includegraphics[width=\textwidth]{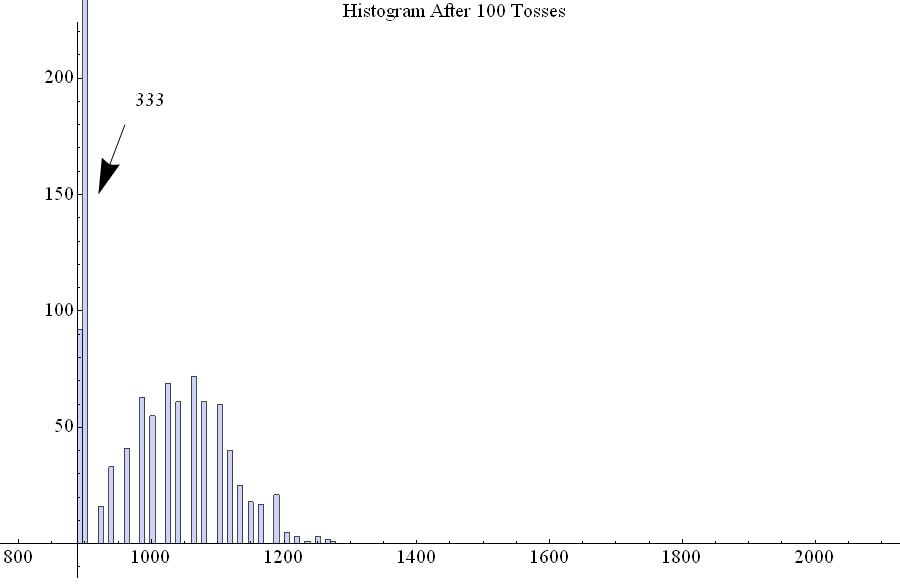}
\end{figure}

\clearpage

\begin{figure} 
  \includegraphics[width=\textwidth]{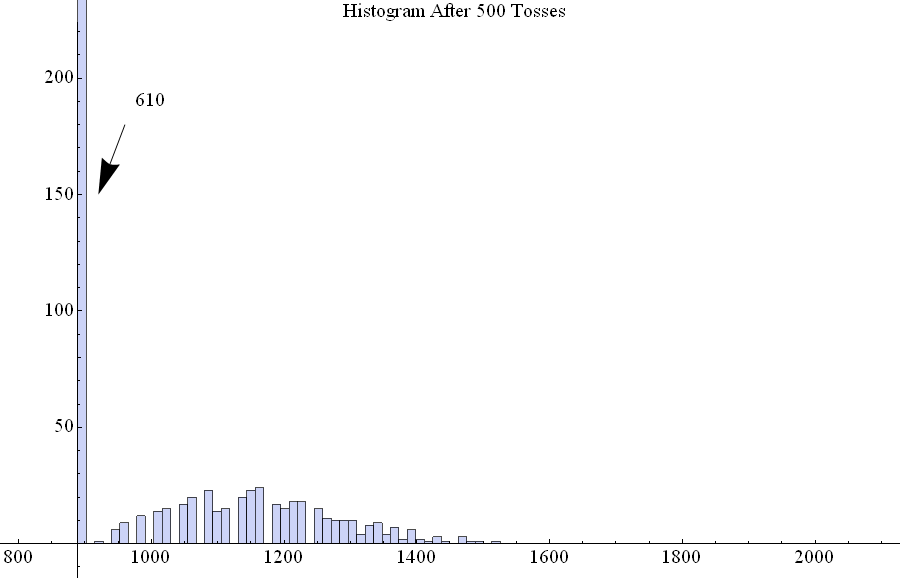}
\end{figure}

\clearpage

\begin{figure} 
  \includegraphics[width=\textwidth]{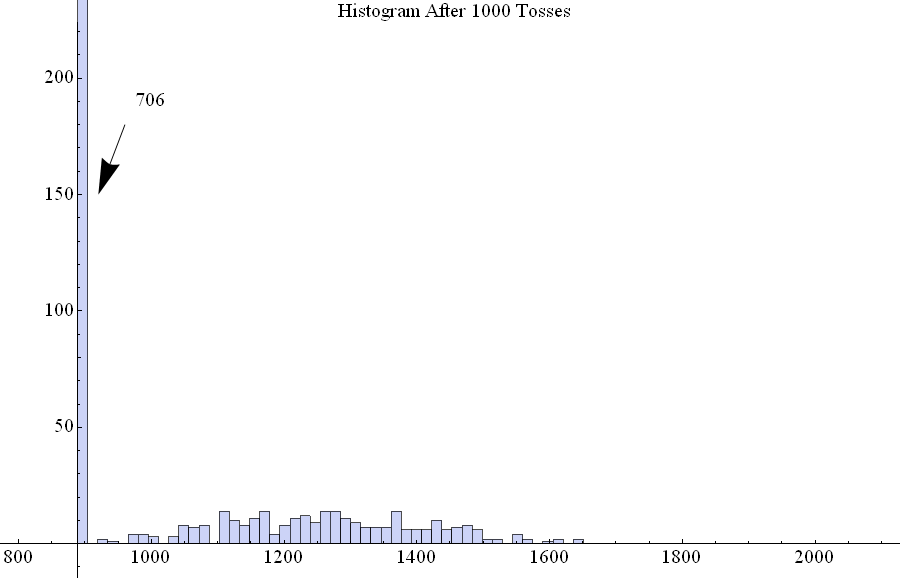}
\end{figure}

\clearpage

\begin{figure} 
  \includegraphics[width=\textwidth]{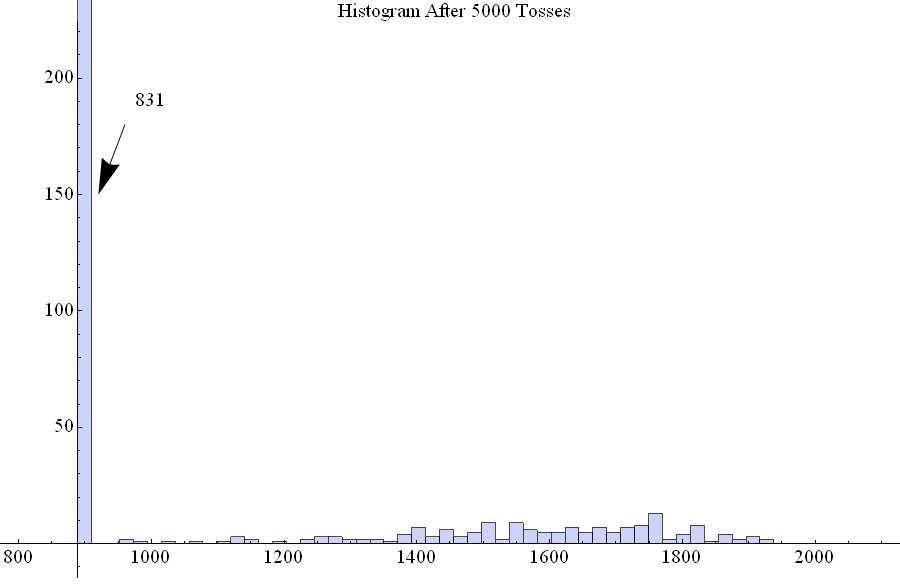}
\end{figure}

\clearpage

\begin{figure} 
  \includegraphics[width=\textwidth]{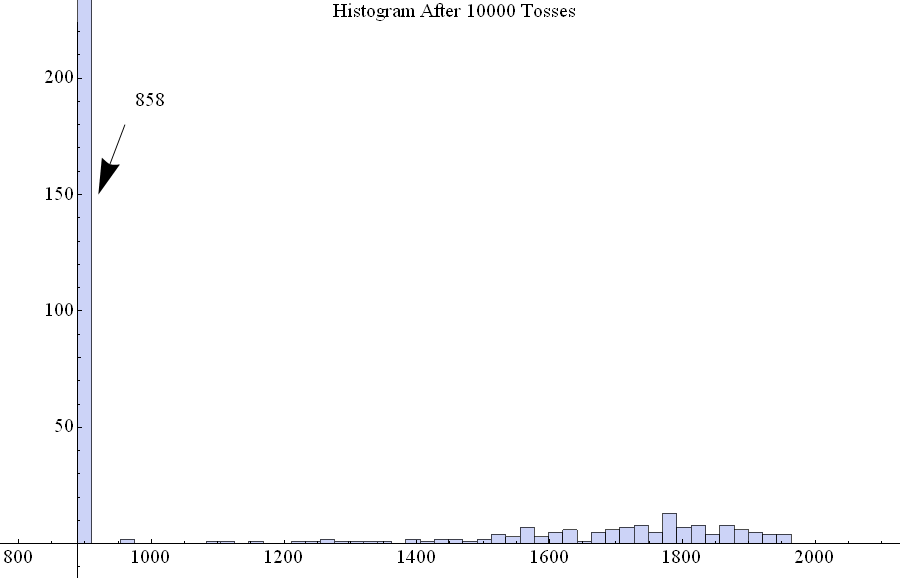}
\end{figure}

\clearpage

\begin{figure} 
  \includegraphics[width=\textwidth]{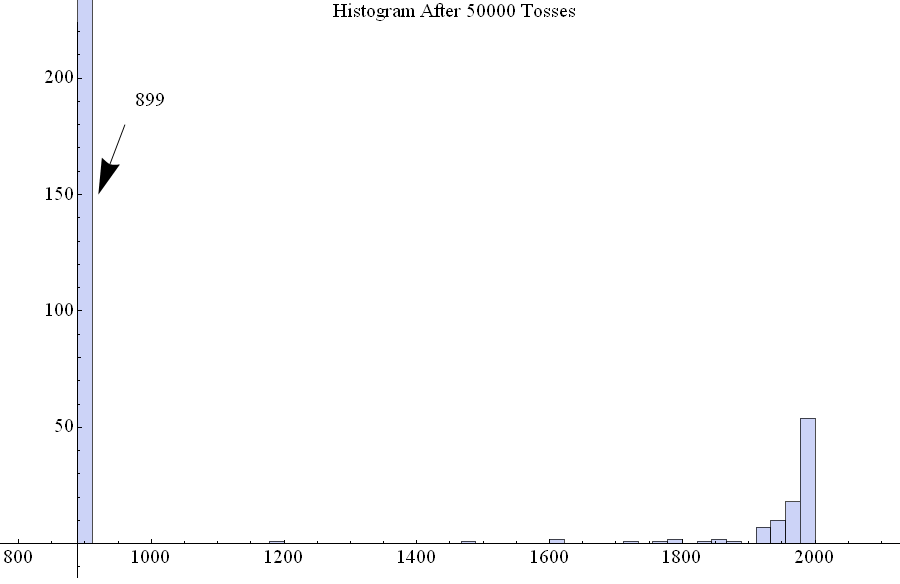}
\end{figure}

\clearpage

\addcontentsline{toc}{section}{References} 

\bibliographystyle{ims}

\end{document}